\documentclass[aps,prd,notitlepage,nofootinbib,superscriptaddress,preprintnumbers,10pt]{revtex4-1}
\usepackage[latin9]{inputenc}
\usepackage{bm}
\usepackage{amsmath}
\usepackage{amssymb}

\makeatletter

\providecommand{\tabularnewline}{\\}

\usepackage{graphicx}
\usepackage{epstopdf}
\usepackage{array}
\def\Mpl{M_{\rm P}}

\makeatother

\begin{document}

\preprint{YITP-17-97, IPMU17-0125}

\title{Horndeski extension of the minimal theory of quasidilaton massive gravity}

\author{Antonio De Felice}
\affiliation{Center for Gravitational Physics, Yukawa Institute for Theoretical Physics, Kyoto University, 606-8502, Kyoto, Japan}
\author{Shinji Mukohyama}
\affiliation{Center for Gravitational Physics, Yukawa Institute for Theoretical Physics, Kyoto University, 606-8502, Kyoto, Japan}
\affiliation{Kavli Institute for the Physics and Mathematics of the Universe (WPI), The University of Tokyo Institutes for Advanced Study, The University of Tokyo, Kashiwa, Chiba 277-8583, Japan}
\author{Michele Oliosi}
\affiliation{Center for Gravitational Physics, Yukawa Institute for Theoretical Physics, Kyoto University, 606-8502, Kyoto, Japan}

\begin{abstract}
 The minimal theory of quasidilaton massive gravity allows for a stable self-accelerating de Sitter solution in a wide range of parameters. On the other hand, in order for the theory to be compatible with local gravity tests, the fifth force due to the quasidilaton scalar needs to be screened at local scales. The present paper thus extends the theory by inclusion of a cubic Horndeski term in a way that (i) respects the quasidilaton global symmetry, that (ii) maintains the physical degrees of freedom in the theory being three, that (iii) can accommodate the Vainshtein screening mechanism and that still (iv) allows for a stable self-accelerating de Sitter solution. After adding the Horndeski term (and a k-essence type nonlinear kinetic term as well) to the precursor action, we switch to the Hamiltonian language and find a complete set of independent constraints. We then construct the minimal theory with three physical degrees of freedom by carefully adding a pair of constraints to the total Hamiltonian of the precursor theory. Switching back to the Lagrangian language, we study cosmological solutions and their stability in the minimal theory. In particular, we show that a self-accelerating de Sitter solution is stable for a wide range of parameters. Furthermore, as in the minimal theory of massive gravity, the propagation speed of the massive gravitational waves in the high momentum limit precisely agrees with the speed of light. 
\end{abstract}

\maketitle

\section{Introduction}

In the context of cosmology, in the most recent years, we have seen the birth of a plethora of new models describing several modifications of gravity. The field of cosmology has seen the expansion of the set of theories which could be responsible and account for the acceleration of the universe measured at large scales~\cite{acceUniv}. Among these theories, we want to mention here the $f(R)$ theories~\cite{DeFelice:2010aj}, the Horndeski's general second-order scalar tensor theories~\cite{Horndeski} and their further
generalizations~\cite{beyondHorndeski}, the generalized Proca theories~\cite{Procas} and finally, massive gravity~\cite{MGhistory,deRham:2010kj,DeFelice:2012mx,DAmico:2012hia,Tolley:2015oxa,Gumrukcuoglu:2011ew,Gumrukcuoglu:2011zh,deRham:2012ew,deRham:2013qqa}, quasidilatonic massive gravity~\cite{DAmico:2012hia,Huang:2012pe,deRham:2014gla,Gumrukcuoglu:2012aa,Gumrukcuoglu:2013nza,D'Amico:2013kya,DeFelice:2013dua,Mukohyama:2014rca,DeFelice:2016tiu,Gabadadze:2014kaa,Kahniashvili:2014wua} and bigravity~\cite{Hassan:2011zd,DeFelice:2013nba,DeFelice:2014nja,DeFelice:2017oym}. 

All these theories consist of generalized Lagrangians which are meant to describe the physics of some propagating degrees of freedom. In particular the complexity has increased from the $\Lambda$-CDM model (two tensor degrees of freedom), Horndeski theory (one scalar, two tensors), Proca theories (one scalar, two vectors, two tensors), massive gravity (one scalar, two vectors, two tensors), quasidilatonic massive gravity (two scalars, two vectors, two tensors), massive bigravity (one scalar, two vectors, four tensors). The phenomenology of each of the mentioned theories tends to be more and more complex as one needs to study the behavior of each new degree of freedom and its effects on already studied phenomena and experiments.

This progression of building more and more complex Lagrangians, has led to a tremendous improvement of understanding about new possibilities of describing low-energy-effective field theories. However still, out of this complexity, it has appeared recently an attempt to have a (simpler) theory which only has two tensor massive modes, i.e.\ a massive gravity with only two degrees of freedom, called the minimal theory of massive gravity~\cite{DeFelice:2015hla}. (See also \cite{Lin:2017oow}.) This feature results from a process of constraining the Lagrangian of massive gravity in order to remove unwanted degrees of freedom. The gain is that homogeneous and isotropic solutions are actually stable solutions of the theory, (a property which was not possible to be implemented in the dRGT theory~\cite{DeFelice:2012mx}), and that a non-trivial phenomenology can be achieved \cite{DeFelice:2015moy} and tested against most recent data \cite{DeFelice:2016ufg}. The price to pay was abandoning Lorentz-invariance, which anyhow is effectively broken only at cosmological scales.

In this same spirit, in order to allow for a scalar-field-induced time-dependence of the fiducial metric \textendash{} which is in the end responsible for the mass of the graviton \textendash{} the minimal theory of quasidilaton massive gravity (hereafter called minimal quasidilaton theory, for brevity) has been introduced \cite{DeFelice:2017wel}. This theory propagates one single scalar (the quasidilaton) and two tensor degrees of freedom. This theory is also the first example of a non-metric field non-trivially coupled with the building-block fields of the minimal theory of massive gravity. This has required a new implementation of the constraints which are necessary to remove the vector modes and the extra-gravity scalar mode.

Our present work, the derivation of the extended version of the minimal quasidilaton theory, which includes a cubic Horndeski Lagrangian for the quasidilaton scalar field, follows steps similar to the original formulation. Furthermore, the Hamiltonian analysis of the quasidilaton sector remains in essence the same as a usual cubic Horndeski theory. We thus also make use of, and present in appendix \ref{sec:cubichorndeski}, the case of the cubic Horndeski Lagrangian, including its Hamiltonian analysis. The main text is organized as follows. In Sec.\ \ref{sec:precursortheory} we present the formulation of the precursor theory, on the basis of both the analysis of the cubic Horndeski theory and the original formulation of the precursor of the minimal quasidilaton theory. Sec.\ \ref{sec:minimaltheory} then describes the addition of constraints leading to the extended minimal theory, given in both Hamiltonian and Lagrangian forms. We describe in Sec.\ \ref{sec:pcase} some features of the Lagrangian in the canonical and k-essence limit of the quasidilaton scalar sector. Further sections are then devoted to the cosmological background solutions (Sec.\ \ref{sec:background}) and the stability on the de Sitter subset of backgrounds (Sec.\ \ref{sec:stability}). Finally, we present our conclusions in Sec.\ \ref{sec:conclusion}. 

The generalization of the minimal quasidilaton theory is a necessary step towards achieving a fully functional model. Indeed, although the original minimal quasidilaton theory is mathematically consistent and allows for a stable self-accelerating cosmological solution in the context of massive gravity, it yet has to implement a screening mechanism at smaller (e.g.\ solar system) scales to be consistent with observations. One of the motivations to generalize the model is thus to add such a screening mechanism. In particular we make use of the Vainshtein mechanism generated by Horndeski terms in the Lagrangian of the quasidilaton. This term offers a screening mechanism at short scales, whereas at large scales, the quasidilaton still leads to IR-modifications of gravity and to massive propagating tensor degrees of freedom. By using a more general set up for the additional scalar field, we have also improved our knowledge of the quasidilaton scalar sector, and this may also allow to ultimately find overarching characteristics of a family of minimal quasidilaton models.

\section{Precursor theory} \label{sec:precursortheory}

The construction of the minimal theory relies on a precursor theory, to which two second-class constraints will be added in order to reduce the number of degrees of freedom to three. In this section we go through the Hamiltonian analysis of this precursor theory. 

\subsection{Precursor Lagrangian}

The Lagrangian density of the precursor theory is given by 
\begin{equation}
\mathcal{L}_\textrm{pre} = \mathcal{L}_{\textrm{E-H}}+\mathcal{L}_{m}+\mathcal{L}_{\sigma}\label{eq:lagrangian_pre}\,,
\end{equation}
where $\mathcal{L}_{\textrm{E-H}}= M^2_\mathrm{P}\sqrt{-g} R[g] /2$ is the Einstein-Hilbert Lagrangian density without cosmological constant, $\mathcal{L}_\sigma$ is the Lagrangian density for the quasidilaton scalar field $\sigma$, and $\mathcal{L}_m$ is the (Lorentz violating) graviton mass term. 

We define the Lagrangian for the quasidilaton scalar field as a cubic Horndeski Lagrangian, using Lagrange multipliers $\theta$, $X$, $\chi$, and $S$, as
\begin{equation}
\mathcal{L}_{\sigma}=\sqrt{-g}\left[F(X,S)+\chi\left(X-\mathfrak{X}\right)+\theta S+g^{\mu\nu}\partial_{\mu}\theta\partial_{\nu}\sigma\right],\label{eq:lagrangian_pre_scalarpart}
\end{equation}
where we have written the canonical kinetic term for the scalar field $\sigma$ as
\begin{equation}
\mathfrak{X}\equiv-\frac{1}{2}g^{\mu\nu}\partial_{\mu}\sigma\partial_{\nu}\sigma\,,\label{eq:canonicalterm}
\end{equation}
and where
\begin{equation}
F(X,S)\equiv P(X)-G(X)S\,,
\end{equation}
in which $P(X)$ and $G(X)$ are sufficiently well-behaved general functions.

The Lagrangian density (\ref{eq:lagrangian_pre_scalarpart}) is equivalent to the usual expression of the cubic Horndeski Lagrangian density
\begin{equation}
 \mathcal{L}'_{\sigma}=\sqrt{-g}F(\mathfrak{X},\Box\sigma)\,,
  \label{eq:lagrangian_standard_horndeski}
\end{equation}
once the e.o.m.\ of $X$, $\chi$ $\theta$, and $S$ are taken into account. The e.o.m.\ of $X$, $\chi$ $\theta$, and $S$ from the Lagrangian density (\ref{eq:lagrangian_pre_scalarpart}) are respectively
\begin{equation}
	\begin{cases}
 	& F_{,X}+\chi=0\,,\\
 	& X-\mathfrak{X}=0\,,\\
 	& S-\Box\sigma=0\,,\\
 	& \theta-G(X)=0\,,
	\end{cases}
\end{equation}
where we have used subscripts after a comma to denote derivatives, for instance, $F_{,X} \equiv \frac{\partial F}{\partial X}$. The system of equations is trivially solved by $\chi=-F_{,X}$, $X=\mathfrak{X}$, $\theta=G(\mathfrak{X})$, and $S=\Box\sigma$, and after substituting this solution to the Lagrangian density (\ref{eq:lagrangian_pre_scalarpart}) one recovers the standard cubic Horndeski Lagrangian density (\ref{eq:lagrangian_standard_horndeski}). (See Appendix~\ref{app:eom-equivalence} for the equivalence of the two systems at the level of equations of motion.) The former is more advantageous for our analysis since by using the Lagrange multipliers one can evade all second or higher derivatives in the Lagrangian density. For this reason, in the rest of the present paper we shall use the Lagrangian density (\ref{eq:lagrangian_pre_scalarpart}).

We use in this work the (3+1) ADM decomposition of the 4-dimensional metric, which necessitates to define the lapse function $N$, the shift vector $N^i$ as well as the spatial 3-dimensional metric $\gamma_{ij}$. These are defined via the line element
\begin{equation}
ds^2 = -N^2 dt^2 + \gamma_{ij}(dx^i + N^i dt)(dx^j + N^j dt)\,,
\end{equation}
where the indices $i,j,\cdots \in \{1,2,3\}$ are used as spatial indices. The 4-dimensional metric $g_{\mu\nu}$ and its inverse $g^{\mu\nu}$ are then given by 
\begin{gather}
g_{00} = -N^2 + \gamma_{ij} N^i N^j\,,\quad g_{0i} = \gamma_{ij} N^i\,,\quad g_{ij} = \gamma_{ij}\,,\\
g^{00} = -\frac{1}{N^2} \,,\quad g^{0i} = \frac{N^i}{N^2}\,,\quad g^{ij} = \gamma^{ij} - \frac{N^i N^j}{N^2}\,,
\end{gather}
where the spatial indices of the shift vector are lowered and raised using the spatial metric $\gamma_{ij}$ and its inverse $\gamma^{ij}$. It is convenient to define 
\begin{equation}
\partial_{\perp}*=\frac{1}{N}\left(\dot{*}-N^{i}\partial_{i}*\right)\,,
\end{equation}
where $\ast$ stands for any field. For example, $\mathfrak{X}$ defined in (\ref{eq:canonicalterm}) is rewritten as 
\begin{equation}
\mathfrak{X}=\frac{1}{2}\left[(\partial_{\perp}\sigma)^{2}-\gamma^{ij}\partial_{i}\sigma\partial_{j}\sigma\right]\,.
\end{equation}

The graviton potential Lagrangian $\mathcal{L}_m$ necessitates the introduction of a fiducial lapse function $M$ and a fiducial spatial metric $\tilde{\gamma}_{ij}$. Out of these we can construct the matrix $\mathcal{K}^i{}_j$ and its inverse $\mathfrak{K}^i{}_j$, such that
\begin{align}
\mathcal{K}^i{}_k\mathcal{K}^k{}_j = \tilde{\gamma}^{ik}\gamma_{kj}\,,\quad \mathcal{K}^i{}_k\mathfrak{K}^k{}_j = \delta^i_j\,.
\end{align}
The traces of $\mathcal{K}^i{}_j$ and $\mathfrak{K}^i{}_j$ as denoted as $\mathcal{K} \equiv \mathcal{K}^i{}_i$ and $\mathfrak{K} \equiv \mathfrak{K}^i{}_i$, and the inverse of the fiducial three dimensional metric is denoted as $\tilde{\gamma}^{ij} = \left(\tilde{\gamma}^{-1}\right)^{ij}$. Using these, the graviton mass Lagrangian density is defined as
\begin{eqnarray}
\mathcal{L}_m & = & \frac{M_{\mathrm{P}}^{2}}{2}\,\sum_{i=0}^{4}\mathcal{L}_{i}\,,\label{eq:lagrangian_gravitonpotential}\\
\mathcal{L}_{0} & = & -m^{2}c_{0}e^{(4+\alpha)\sigma/M_{\mathrm{P}}}\,\sqrt{\tilde{\gamma}}\,M\,,\label{eqn:def-S0}\\
\mathcal{L}_{1} & = & -m^{2}c_{1}e^{3\sigma/M_{\mathrm{P}}}\,\sqrt{\tilde{\gamma}}\,(N+Me^{\alpha\sigma/M_{\mathrm{P}}}\mathcal{K})\,,\\
\mathcal{L}_{2} & = & -m^{2}c_{2}e^{2\sigma/M_{\mathrm{P}}}\,\sqrt{\tilde{\gamma}}\,\left[N\mathcal{K}+\frac{1}{2}Me^{\alpha\sigma/M_{\mathrm{P}}}(\mathcal{K}^{2}-\mathcal{K}^{i}{}_{j}\mathcal{K}^{j}{}_{i})\right]\,,\\
\mathcal{L}_{3} & = & -m^{2}c_{3}e^{\sigma/M_{\mathrm{P}}}\sqrt{\gamma}\,(N\,\mathfrak{K}+Me^{\alpha\sigma/M_{\mathrm{P}}})\,,\\
\mathcal{L}_{4} & = & -m^{2}c_{4}\sqrt{\gamma}N\,. \label{eqn:def-S4}
\end{eqnarray}
The contribution from $\mathcal{L}_{4}$ is effectively a cosmological constant term; it is therefore not necessary to include a cosmological constant in $\mathcal{L}_\textrm{E-H}$. The graviton mass term is linear in the lapses and does not depend on the shift variables. This is a consequence of the Lorentz violations in the gravity sector, and is in a sharp contrast to the dRGT theory, whose graviton mass term is nonlinear in the lapse function and depends on the shift vector. We have introduced 7 parameters, $\{m,c_0,c_1,c_2,c_3,c_4,\alpha\}$, including $c_4$ playing the role of the cosmological constant. The parameters $c_i$ are used in dRGT massive gravity \cite{deRham:2010kj} to parametrize the most general Lorentz invariant ghost-free mass term. Here, both the use of the ADM frame and the presence of the parameter $\alpha$ signify a breaking of Lorentz symmetry, which is however confined to the gravity sector and is suppressed by the small parameter $m \sim H_{\textrm{today}}$. 

The Lagrangian (\ref{eq:lagrangian_gravitonpotential}) is constructed as to be invariant under the quasidilaton global symmetry transformation. The quasidilaton global symmetry transformation is defined by their action on the St\"{u}ckelberg fields $\left(\phi^0,\phi^1,\phi^2,\phi^3\right)$, introduced in order to recover general covariance at the level of the action. In the covariant picture, the transformation rule is 
\begin{equation}
\sigma \rightarrow \sigma + \sigma_0\,,\quad \phi^a \rightarrow \phi^a e^{-\sigma_0/M_\mathrm{P}}\,,\quad \phi^0 \rightarrow \phi^0 e^{-(1+\alpha)\sigma_0} \,,
\end{equation}
where $\alpha$ is the constant in (\ref{eqn:def-S0})-(\ref{eqn:def-S4}). Out of the St\"{u}ckelberg fields and the metric, one can build the combinations
\begin{equation}
\boldsymbol{N} \equiv \frac{1}{\sqrt{-g^{\mu\nu}\partial_\mu\phi^0\partial_\nu\phi^0}}\,,\quad \boldsymbol{N}^p \equiv \boldsymbol{N}^2g^{\mu\nu}\partial_\mu\phi^p\partial_\nu\phi^0\,,\quad {\boldsymbol{\gamma}}^{pq}\equiv g^{\mu\nu}\partial_\mu\phi^p\partial_\nu\phi^q + \frac{\boldsymbol{N}^p\boldsymbol{N}^q}{\boldsymbol{N}^2}\,.
\end{equation}
Furthermore, the fiducial quantities in the covariant formulation, $\boldsymbol{M}$ and $\tilde{\boldsymbol{\gamma}}_{pq}$, are defined as functions of the St\"{u}ckelberg variables. One further requires invariance under $SO(3)$ and translations in the field space spanned by the $\phi^p$, and invariance under shifts of $\phi^0$, all of which allows one to fix the fiducial variables to $\boldsymbol{M}=1$ and $\tilde{\boldsymbol{\gamma}}_{pq} = \delta_{pq}$, and which implies that we actually do not need to introduce the fiducial lapse. The Lagrangian is then built so that the quasidilaton global symmetry is maintained. Once the unitary gauge $\phi^\mu = x^\mu$  is chosen, the St\"{u}ckelberg fields disappear from the action and the combinations $\boldsymbol{N}$, $\boldsymbol{N}^p$ and $\boldsymbol{\gamma}^{pq}$ reduce to the ADM variables.

The quasidilaton global symmetry is what gives the quasidilaton scalar field its interest in the context of massive gravity, as its dynamics can for instance and among others induce an exponential scaling of the fiducial metric. In the expressions that follow we will keep both $M=1$ and $\tilde{\gamma}_{ij}=\delta_{ij}$ whenever they appear, as they are useful as a check for the expressions in which they appear.

\subsection{Primary constraints, Hamiltonian and consistency conditions}

We consider $\{\gamma_{(ij)},\sigma,X,\chi,\theta,S\}$ and their conjugate momenta as 22 canonical variables, and $\{N,N^{i}\}$ as Lagrange multipliers, as these only appear linearly in the action. Upon calculating
the conjugate momenta, we get 
\begin{gather}
\pi^{ij}\equiv\frac{M_{\mathrm{P}}^{2}}{2}\sqrt{\gamma}\left(K^{ij}-K\gamma^{ij}\right)\,,\quad\pi_{\sigma}\equiv-\sqrt{\gamma}(\chi\partial_{\perp}\sigma+\partial_{\perp}\theta)\,,\quad\pi_{\theta}\equiv-\sqrt{\gamma}(\partial_{\perp}\sigma)\,,\label{eq:momenta_pre}\\
\pi_{\chi}=0\,, \quad \pi_{S}=0\,,\quad\pi_{X}=0\,,
\end{gather}
where 
\begin{equation}
K_{ij}=\frac{1}{2N}\left(\dot{\gamma}_{ij}-\mathcal{D}_{i}N_{j}-\mathcal{D}_{j}N_{i}\right)\,.
\end{equation}
As an intermediate step before computing the Hamiltonian, we invert relations (\ref{eq:momenta_pre}) as 
\begin{align}
\dot{\gamma}_{ij}=2NK_{ij}(\pi^{kl})+\mathcal{D}_{i}N_{j}+\mathcal{D}_{j}N_{i}\,,\quad
\dot{\sigma}=-N\tilde{\pi}_{\theta}+N^{i}\partial_{i}\sigma\,,\quad
\dot{\theta}=N(\chi\tilde{\pi}_{\theta}-\tilde{\pi}_{\sigma})+N^{i}\partial_{i}\theta\,.
\end{align}
We have found it useful to define the tilded momenta as three dimensional scalars, i.e., for instance, $\tilde{\pi}_\theta \equiv \frac{\pi_\theta}{\sqrt{\gamma}}$. In addition to previous relations, the primary constraints related to $X$, $\chi$, and $S$ are defined as 
\begin{equation}
0=P_{X}\equiv\pi_{X}\,,\quad0=P_{\chi}\equiv\pi_{\chi}\,,\quad0=P_{S}\equiv\pi_{S}\,.
\end{equation}
We can thus again define the Hamiltonian with all primary constraints as
\begin{equation}
\bar{H}_{\mathrm{pre}}^{(1)}=\int d^{3}x\left[-N\mathcal{R}_{0}-N^{i}\mathcal{R}_{i}+\frac{M_{\mathrm{P}}^{2}}{2}m^{2}M\mathcal{H}_{1}+ \xi_X P_X + \xi_\chi P_\chi + \xi_S P_S\right],
\end{equation}
where 
\begin{eqnarray*}
\mathcal{R}_{0} & = & \mathcal{R}_{0}^{\mathrm{GR}}-\frac{M_{\mathrm{P}}^{2}}{2}m^{2}\mathcal{H}_{0}\,,\\
\mathcal{R}_{0}^{\mathrm{GR}} & = & \frac{M_{\mathrm{P}}^{2}}{2}\sqrt{\gamma}\,R[\gamma]-\frac{2}{M_{\mathrm{P}}^{2}}\frac{1}{\sqrt{\gamma}}\left(\gamma_{il}\gamma_{jk}-\frac{1}{2}\gamma_{ij}\gamma_{kl}\right)\pi^{ij}\pi^{kl}-\mathcal{H}_{\sigma}\,,\\
\mathcal{H}_{\sigma} & = & \sqrt{\gamma} \left[\frac{\chi}{2}\frac{\pi_\theta^2}{\sqrt{\gamma}^2}-\frac{\pi_\theta\pi_\sigma}{\sqrt{\gamma}^2}-F-\chi\left(X+\frac{1}{2}\gamma^{ij}\partial_{i}\sigma\partial_{j}\sigma\right)-\theta S -\gamma^{ij}\partial_i\sigma\partial_j\theta\right]\,,\\
\mathcal{R}_{i} & = & 2\sqrt{\gamma}\gamma_{ik}\mathcal{D}_{j}\left(\frac{\pi^{kj}}{\sqrt{\gamma}}\right)-\pi_\sigma\partial_{i}\sigma-\pi_\theta\partial_{i}\theta\,,\\
\mathcal{H}_{0} & = & \mathcal{H}_{0}(\sigma, \gamma_{ij}) = \sqrt{\tilde{\gamma}}\left(c_{1}e^{3\sigma/M_{\mathrm{P}}}+c_{2}e^{2\sigma/M_{\mathrm{P}}}\,\mathcal{K}\right)+\sqrt{\gamma}(c_{3}e^{\sigma/M_{\mathrm{P}}}\,\mathfrak{K}+c_{4}) \,,\\
\mathcal{H}_{1} & = & \mathcal{H}_{1}(\sigma, \gamma_{ij}) = e^{\alpha\sigma/M_{\mathrm{P}}}\left\{ \sqrt{\tilde{\gamma}}\left[c_{0}e^{4\sigma/M_{\mathrm{P}}}+c_{1}e^{3\sigma/M_{\mathrm{P}}}\mathcal{K}+\frac{c_{2}}{2}e^{2\sigma/M_{\mathrm{P}}}\,(\mathcal{K}^2-\mathcal{K}^i{}_j\mathcal{K}^j{}_i)\right]+c_{3}\sqrt{\gamma}e^{\sigma/M_{\mathrm{P}}}\right\} \,.
\end{eqnarray*}

In the cubic Horndeski case without graviton mass terms, it is found that if we want to have the momentum constraint as a generator for translations, we need to modify it so as to include $\pi_X$, $\pi_\chi$, and $\pi_S$. We thus define 
\begin{align}
\tilde{\mathcal{R}}_{i} & \equiv2\sqrt{\gamma}\gamma_{ik}\mathcal{D}_{j}\tilde{\pi}^{kj}-\pi_{\sigma}\partial_{i}\sigma-\pi_{X}\partial_{i}X-\pi_{\chi}\partial_{i}\chi-\pi_{S}\partial_{i}S-\pi_{\theta}\partial_{i}\theta\,.\label{eq:defmomentumconstrainttilde}
\end{align}
Clearly, this momentum constraint is equivalent to the original one when restricted to the constraint surface. Although in the presence of graviton potential terms in the action this constraint is not a generator of spatial diffeomorphisms anymore, it still proves useful in order to simplify the constraint algebra and the Hamiltonian structure. We turn to analyzing the primary constraint algebra, which we summarize in table \ref{table:primaryconstraintalgebra_pre}.

\begin{table}[ht]
	\begin{tabular}{c|ccccc}
		\{$\downarrow$,$\rightarrow$\} & $\mathcal{R}_0$ & $\tilde{\mathcal{R}}_i$ & $P_X$ & $P_\chi$ & $P_S$ \\
		\hline $\mathcal{R}_0$ & $0$ & $\not\approx0$ & $-(\chi+F,_X)$ & $-(X-\mathfrak{X})$ & $-(\theta-G(X))$ \\
		$\tilde{\mathcal{R}}_i$ & \multicolumn{1}{c}{}  & $0$ & $0$ & $0$ & $0$ \\
		$P_X$ & \multicolumn{2}{c}{} & $0$ & $0$ & $0$\\
		$P_\chi$ & \multicolumn{3}{c}{} & $0$ & $0$ \\
		 $P_S$ & \multicolumn{4}{c}{} & $0$ 
	\end{tabular}\caption{Primary constraint algebra of the precursor theory. Dirac $\delta$-functions were omitted in the entries. When $\not\approx 0$ is indicated, the entry may formally include not only Dirac $\delta$-functions but also derivatives of Dirac $\delta$-functions.}\label{table:primaryconstraintalgebra_pre}
\end{table}
We give some results in their integral form,
\begin{align}
\{ P_\star,\tilde{\mathcal{R}}_i [f^i] \} & = - \int d^3x \sqrt{\gamma}\,\mathcal{D}_i\!\left(P_\star f^i\right) \approx 0\, ,\label{eq:poisson_primaryWmomentum}\\
\{ \mathcal{R}_0 [\phi_2], \mathcal{R}_0[\phi_2] \} & = \int d^3x \mathcal{R}_i \left(\phi_1\mathcal{D}^i \phi_2-\phi_2\mathcal{D}^i \phi_1\right) \approx 0\,,\\
\{ \tilde{\mathcal{R}}_i [f^i], \tilde{\mathcal{R}}_j[g^j] \} & = \int d^3x \tilde{\mathcal{R}}_i \left(g^j\mathcal{D}_j f^i -f^j\mathcal{D}_jg^i\right) \approx 0\,.
\end{align}
One can observe that the Hamiltonian and momentum constraints obey the usual algebra. In equation (\ref{eq:poisson_primaryWmomentum}) the symbol $\star$ stands for any of $X$, $\chi$, and $S$. 

Consistency of the primary constraints $P_X$, $P_\chi$, and $P_S$ with the time evolution of the system yields the following conditions, which cannot be solved for Lagrange multipliers (unless one sets N to be zero, which is unphysical):
\begin{align}
\dot{P}_X  \equiv \sqrt{\gamma}\{\pi_X,\bar{H}_\mathrm{pre}^{(1)}\} & = N \sqrt{\gamma} (\chi+F,_X) \approx 0 \\
\dot{P}_\chi \equiv \sqrt{\gamma}\{\pi_\chi,\bar{H}_\mathrm{pre}^{(1)}\} & = N \sqrt{\gamma} (X-\mathfrak{X}_H) \approx 0 \\
\dot{P}_S   \equiv \sqrt{\gamma}\{\pi_S,\bar{H}_\mathrm{pre}^{(1)}\} & =  N \sqrt{\gamma} (\theta-G(X)) \approx 0 \,,
\end{align}
where
\begin{equation}
\mathfrak{X}_H \equiv \frac{1}{2}\left(\tilde{\pi}_\theta^2 - \gamma^{ij}\partial_i\sigma\partial_j\sigma\right).
\end{equation}
We thus use these conditions to define the secondary constraints
\begin{eqnarray}
S_X (X, \chi, S) &\equiv& \chi+F,_X\,, \label{eq:constraint_secondary_X}\\
S_\chi (\gamma, \sigma, X, \pi_\theta) &\equiv& X-\mathfrak{X}_H\,,\label{eq:constraint_secondary_chi}\\
S_S (\theta,X) &\equiv& \theta-G(X) \label{eq:constraint_secondary_S}\,.
\end{eqnarray}
Other secondary constraints stem from the time evolution of $\mathcal{R}_{0}$ and $\tilde{\mathcal{R}}_{i}$. We have that in general
\begin{align}
\{\mathcal{R}_{0},\tilde{\mathcal{R}}_{i}\} \not\approx 0\,,\\
\{\tilde{\mathcal{R}}_{i},\tilde{\mathcal{R}}_{j}\} \approx 0\,,\\
\{\mathcal{R}_{0},\mathcal{R}_{0}\} \approx 0\,.
\end{align}
As only the consistency condition for $\mathcal{R}_{0}$ and one of the three consistency conditions for $\mathcal{R}_{i}$ may be solved for Lagrange multipliers (in this case $N$ and one of the $N^i$) we need to impose the two additional secondary constraints, which we call $\tilde{\mathcal{C}}_{\tau}$.

\subsection{Secondary Hamiltonian, constraint algebra, and tertiary constraints}

We define the secondary Hamiltonian with all primary and secondary constraints
\begin{equation}
\bar{H}_{\mathrm{pre}}^{(2)}=\int d^{3}x\,[-N\mathcal{R}_{0}-N^{i}\tilde{\mathcal{R}}_{i}+\frac{M_{\mathrm{P}}^{2}}{2}m^{2}M\mathcal{H}_{1}+\xi_{X}P_{X}+\xi_{\chi}P_{\chi}+\xi_{S}P_{S}+\sqrt{\gamma}\left(\lambda_{X}S_{X}+\lambda_{\chi}S_{\chi}+\lambda_{S}S_{S}\right)+\tilde{\lambda}^{\tau}\tilde{\mathcal{C}}_{\tau}]\,.
\end{equation}

The secondary constraint algebra is summarized in the table \ref{table_algebrasecondary_pre}, where $T_{\chi}$ and $T_{S}$ stand for 
\begin{align}
T_{\chi} & =\frac{2}{M_{\textrm{P}}^{2}}\,\tilde{\pi}^{ij}\left[\gamma_{ij}\mathfrak{X}_H-\mathcal{D}_{i}\sigma\mathcal{D}_{j}\sigma\right]-\tilde{\pi}_{\theta}\left[S-\gamma^{ij}\mathcal{D}_{i}\mathcal{D}_{j}\sigma\right]-\gamma^{ij}\mathcal{D}_{j}\sigma\mathcal{D}_{i}\tilde{\pi}_{\theta}\,,\label{eq:bracket_tertiary_chi}\\
T_{S} & =\chi\tilde{\pi}_{\theta} - \tilde{\pi}_{\sigma}\label{eq:bracket_tertiary_S}\,.
\end{align}

\begin{table}[ht]
\begin{tabular}{c|cccccccccc}
\{$\downarrow$,$\rightarrow$\} & $\mathcal{H}_{1}$  & $\mathcal{R}_{0}$  & $\tilde{\mathcal{R}}_{i}$  & $\tilde{\mathcal{C}}_{\tau}$   & $P_{X}$  & $P_{\chi}$  & $P_{S}$  & $S_{X}$  & $S_{\chi}$  & $S_{S}$ \tabularnewline
\hline 
$\mathcal{H}_{1}$ & $0$  & $\not\approx 0$  & $\not\approx 0$  & $0$ & $0$  & $0$  & $0$  & $0$  & $0$  & $0$ \tabularnewline
$\mathcal{R}_{0}$  &  \multicolumn{1}{c}{} & $0$  & $\not\approx 0$  & $\not\approx 0$ & $0$  & $0$  & $0$  & $0$  & $T_{\chi}$  & $T_{S}$ \tabularnewline
$\tilde{\mathcal{R}}_{i}$ & \multicolumn{2}{c}{}  & $0$  & $\not\approx 0$ & $0$  & $0$  & $0$  & $0$  & $0$  & $0$ \tabularnewline
$\tilde{\mathcal{C}}_{\tau}$ & \multicolumn{3}{c}{} & $0$ & $0$  & $0$  & $0$  & $0$  & $0$  & $0$ \tabularnewline
$P_{X}$ & \multicolumn{4}{c}{}  & $0$  & $0$  & $0$  & $-F_{,XX}$  & $-1$  & $G_{,X}$ \tabularnewline
$P_{\chi}$ & \multicolumn{5}{c}{}  & $0$  & $0$  & $-1$  & $0$ & $0$ \tabularnewline
$P_{S}$ & \multicolumn{6}{c}{}  & $0$  & $G_{,X}$  & $0$  & $0$\tabularnewline
$S_{X}$ & \multicolumn{7}{c}{} & $0$  & $0$  & $0$ \tabularnewline
$S_{\chi}$ & \multicolumn{8}{c}{}  & $0$  & $\pi_{\theta}/\sqrt{\gamma}$ \tabularnewline
$S_{S}$ & \multicolumn{9}{c}{}  & $0$
\end{tabular}\caption{Secondary constraint algebra of the precursor theory. Dirac $\delta$-functions were omitted in the entries. When $\not\approx 0$ is indicated, the entry may formally include not only Dirac $\delta$-functions but also derivatives of Dirac $\delta$-functions.}
\label{table_algebrasecondary_pre} 
\end{table}

It can be readily checked that the constraints $P_X$, $P_{\chi}$,
$P_S$, $S_X$, $S_{\chi}$, and $S_S$ all commute with the non-vanishing Poisson bracket $\{\tilde{\mathcal{R}}_i(x), \mathcal{R}_0 (y)\}$. 

The consistency conditions yield the following equations 
\begin{align}
\dot{P}_{\chi} \approx 0 \approx \dot{P}_{S} & :\lambda_{X}\approx 0\\
\dot{P}_{X} \approx 0 & :\lambda_{\chi}-\lambda_{S}G_{,X}\approx 0\label{consistencyPX}\\
\dot{S}_{\chi} \approx 0 & :\xi_{X}+\lambda_{S}\tilde{\pi}_{\theta}\approx 0\label{consistencySchi}\\
\dot{S}_{X} \approx 0 & :\xi_{X}F_{,XX}+\xi_{\chi}-\xi_{S}G_{,X}\approx 0\\
\dot{S}_{S} \approx 0 & :N(T_{S}+G_{,X}T_{\chi})+\xi_{X}G_{,X}+\lambda_{\chi}\tilde{\pi}_{\theta}\approx 0\label{consistencySS}\\
\dot{\tilde{\mathcal{R}}}_{0}\approx 0 & :\lambda_{S}(T_{S}+G_{,X}T_{\chi})\approx 0\,.
\end{align}
By plugging Eqs.\ (\ref{consistencyPX}) and (\ref{consistencySchi}) into
Eq.\ (\ref{consistencySS}), we obtain 
\begin{equation}
N\,(T_{S}+G_{,X}T_{\chi})=0.
\end{equation}

As a consequence, since setting the lapse to zero would be unphysical,
we need to impose the tertiary constraint $T\approx 0$, where
\begin{equation}
T\equiv T_{S}+G_{,X}T_{\chi}=\chi\tilde{\pi}_{\theta}-\tilde{\pi}_{\sigma}-G_{,X}\left[\frac{2}{M_{\textrm{P}}^{2}}\,\tilde{\pi}^{ij}\left(\mathcal{D}_{i}\sigma\mathcal{D}_{j}\sigma+\gamma_{ij}\mathfrak{X}_H\right)+\tilde{\pi}_{\theta}\left(S-\gamma^{ij}\mathcal{D}_{i}\mathcal{D}_{j}\sigma\right)+\gamma^{ij}\mathcal{D}_{j}\sigma\mathcal{D}_{i}\tilde{\pi}_{\theta}\right] .
\end{equation}

Due to the Lorentz-symmetry-breaking mass term, the Hamiltonian constraint should end up as a second-class constraint and not be a generator of time diffeomorphisms. Nevertheless, we can relevantly simplify the algebra of constraints by defining the new combinations
\begin{align}
\tilde{\mathcal{R}}_{0} & =\mathcal{R}_{0}-T_{\chi}(P_{X}-F_{,XX}P_{\chi})\,,\\
\tilde{P}_{X} & =P_{X}-F_{,XX}P_{\chi}\,,\\
\tilde{P}_{S} & =P_{S}+G_{,X}P_{\chi}\,,\\
\tilde{S}_{S} & =S_{S}+G_{,X}S_{\chi}-\tilde{\pi}_{\theta}\tilde{P}_{X}\,,\\
\tilde{Q} & =\{T,\tilde{\mathcal{R}}_{0}\}\,,\\
\bar{\mathcal{R}}_{0} & =\tilde{\mathcal{R}}_{0}+\frac{\tilde{Q}}{B}\tilde{S}_{S}\,.
\end{align}
It can be easily verified that $\tilde{P}_S$ is a first-class constraint. For brevity, we do not write an explicit expression for $\tilde{Q}$, which however does not trivially vanish and is proportional to a Dirac $\delta$-function, without derivatives, and is thus a local expression.  In the absence of graviton Lorentz-symmetry-breaking potential terms in the action, the new Hamiltonian becomes the generator of time diffeomorphisms.

The resulting tertiary constraint algebra is given in Table \ref{table_algebratertiary_pre}, where we have omitted the first-class constraint ${\tilde P}_S$. Table entries $A$, $D$, and $B$ do not trivially vanish and are proportional to Dirac $\delta$-functions, and are thus strictly local expressions. Table entries $Q_1$, $Q_2$, $Q_\chi$, and $Q_T$ do not trivially vanish and include derivatives of Dirac $\delta$-functions. As the explicit expressions are not directly needed for the analysis, we do not give these here explicitly. The reader may find some components in appendix \ref{sec:cubichorndeski}, as most of these expressions coincide with the simpler cubic Horndeski case.

\begin{table}[ht]
\begin{tabular}{c|cccccccccc} 
\{$\downarrow$,$\rightarrow$\} & $\mathcal{H}_{1}$  & $\bar{\mathcal{R}}_{0}$  & $\tilde{\mathcal{R}}_{i}$  & $\tilde{\mathcal{C}}_{\tau}$ & $\tilde{P}_{X}$  & $P_{\chi}$  & $S_{X}$  & $S_{\chi}$  & $\tilde{S}_{S}$ & $T$ \tabularnewline
\hline 
$\mathcal{H}_{1}$  & $0$  & $\not \approx 0$  & $\not\approx 0$  & $0$  & $0$  & $0$  & $0$  & $0$  & $0$  & $0$ \tabularnewline
$\bar{\mathcal{R}}_{0}$  & \multicolumn{1}{c}{} & $0$  & $\not\approx 0$  & $\not\approx 0$  & $0$  & $0$  & $0$  & $0$  & $0$  & $0$ \tabularnewline
$\tilde{\mathcal{R}}_{i}$  & \multicolumn{2}{c}{} & $0$  & $\not\approx 0$  & $0$  & $0$  & $0$  & $0$  & $0$  & $0$ \tabularnewline
$\tilde{\mathcal{C}}_{\tau}$  & \multicolumn{3}{c}{} & $0$ & $0$  & $0$  & $0$  & $0$  & $0$  & $-Q_{\tau}$ \tabularnewline 
$\tilde{P}_{X}$  & \multicolumn{4}{c}{} & $0$  & $0$  & $0$  & $-1$  & $0$  & $A$ \tabularnewline
$P_{\chi}$  & \multicolumn{5}{c}{} & $0$  & $-1$  & $0$  & $0$  & $D$ \tabularnewline
$S_{X}$  & \multicolumn{6}{c}{} & $0$  & $0$  & $0$  & $0$ \tabularnewline
$S_{\chi}$  & \multicolumn{7}{c}{} & $0$  & $0$  & $-Q_{\chi}$ \tabularnewline
$\tilde{S}_{S}$  & \multicolumn{8}{c}{} & $0$  & $B$ \tabularnewline
$T$  & \multicolumn{9}{c}{} & $Q_{T}$
\end{tabular}\caption{Tertiary constraint algebra of the precursor theory. The first-class constraint $\tilde{P}_S$ was omitted. Dirac $\delta$-functions were omitted in the entries, excepted for $Q_1$, $Q_2$, $Q_\chi$, and $Q_T$, which formally include derivatives of $\delta$-functions. When $\not\approx 0$ is indicated, the entry may formally include not only Dirac $\delta$-functions but also derivatives of Dirac $\delta$-functions.}
\label{table_algebratertiary_pre} 
\end{table}

At this point there are 12 second-class and 1 first-class ($\tilde{P}_S$) constraints. This reduces the 22 phase space degrees of freedom (6 metric components, 1 quasidilaton, 4 scalar fields Horndeski-Lagrange multipliers) to 8 phase space degrees of freedom, i.e.\ 4 degrees of freedom, as expected. As we will see, the minimal theory is defined by adding two supplementary second-class constraint, thus effectively reducing the number of degrees of freedom to 3.

\section{Minimal theory} \label{sec:minimaltheory}

\subsection{Hamiltonian of the minimal theory }
In the previous section we have obtained the total Hamiltonian of the precursor theory, after going through a complete Hamiltonian analysis. Now, one can define the Hamiltonian of the minimal quasidilaton theory as 
\begin{eqnarray}
  \bar{H}^{(T)}=\int d^{3}x&&\left[-N\tilde{\mathcal{R}}_{0}-N^{i}\tilde{\mathcal{R}}_{i}+\frac{M_{\mathrm{P}}^{2}}{2}m^{2}M\mathcal{H}_{1}+\xi_{X}P_{X}+\xi_{\chi}P_{\chi}+\xi_{S}P_{S}+\sqrt{\gamma}\left(\lambda_{X}S_{X}+\lambda_{\chi}S_{\chi}+\lambda_{S}S_{S}+\lambda_T\tilde{T}\right)\right.\nonumber\\
 &&{}+\left.\frac{M_{\mathrm{P}}^{2}}{2}\left(\lambda^{i}\mathcal{C}_{i}+\lambda\mathcal{C}_{0}\right)\right]\,,
\end{eqnarray}
where $\mathcal{C}_{i}$ ($i=1,2,3$) are defined as follows 
\begin{equation}
\{\tilde{\mathcal{R}}_{i}^{\mathrm{GR}},H_{1}\}\approx\frac{M_{\mathrm{P}}^{2}}{2}\,\mathcal{C}_{i}\,,\quad\{\tilde{\mathcal{R}}_{0}^{\mathrm{GR}},H_{1}\}\approx\frac{M_{\mathrm{P}}^{2}}{2}\,\mathcal{C}_{0}\,,
\end{equation}
where 
\begin{equation}
H_{1}=\frac{M_{\mathrm{P}}^{2}}{2}\,m^{2}\int d^{3}xM\mathcal{H}_{1}\,.
\end{equation}
We write here the explicit expression of the new constraints. These are 
\begin{eqnarray}
\mathcal{C}_{0} & = & m^{2}M\left[\frac{1}{M_{\mathrm{P}}^{2}}\left(\gamma_{ik}\gamma_{jl}- \frac{1}{2}\gamma_{ij}\gamma_{kl}\right)\Theta^{ij}\pi^{kl}-\frac{\partial\mathcal{H}_{1}}{\partial\sigma}\tilde{\pi}_{\theta}\nonumber\right]\,,\nonumber \\
\mathcal{C}_{i} & = & m^{2}\left[-\frac{1}{2}\sqrt{\gamma}\mathcal{D}_{j}\!\left(M\Theta^{jk}\gamma_{ki}\right)+M\frac{\partial\mathcal{H}_{1}}{\partial\sigma}\partial_{i}\sigma\right]\,.
\end{eqnarray}
Here we have defined 
\begin{eqnarray}
\Theta^{ij} & = & e^{\alpha\sigma/M_{\mathrm{P}}}\left\{ \frac{\sqrt{\tilde{\gamma}}}{\sqrt{\gamma}}\left[\left(c_{1}e^{3\sigma/M_{\mathrm{P}}}+c_{2}e^{2\sigma/M_{\mathrm{P}}}\mathcal{K}\right)\left(\mathcal{K}^i{}_k\gamma^{kj}+\gamma^{ik}\mathcal{K}^j{}_k\right)-2c_{2}e^{2\sigma/M_{\mathrm{P}}}\tilde{\gamma}^{ij}\right]+2c_{3}e^{\sigma/M_{\mathrm{P}}}\gamma^{ij}\right\} \,,\\
\frac{\partial\mathcal{H}_{1}}{\partial\sigma} & = & \frac{e^{\alpha\sigma/M_{\mathrm{P}}}}{M_{\mathrm{P}}}\left\{ \sqrt{\tilde{\gamma}}\left[(4+\alpha)c_{0}e^{4\sigma/M_{\mathrm{P}}}+(3+\alpha)c_{1}e^{3\sigma/M_{\mathrm{P}}}\mathcal{K}\right.\right.\nonumber \\
 &  & \left.\left.+\frac{2+\alpha}{2}c_{2}e^{2\sigma/M_{\mathrm{P}}}\,(\mathcal{K}^{2}-\mathcal{K}^{i}{}_{j}\mathcal{K}^{j}{}_{i})\right]+(1+\alpha)c_{3}\sqrt{\gamma}e^{\sigma/M_{\mathrm{P}}}\right\} \,,
\end{eqnarray}
and we have combined the tertiary constraint with the secondary constraints $S_{\chi}$ and $S_X$ as
\begin{equation}
\tilde{T}=T-G_{,X}\frac{2}{M_{\mathrm{P}}^{2}}\tilde{\pi} S_{\chi} - \tilde{\pi}_\theta S_X\,,\label{eq:newOldT}
\end{equation}
so that no term cubic in the momenta is present. Introduction of the combination (\ref{eq:newOldT}) will simplify the analysis when we invert the momenta in order to find the Lagrangian. Moreover, the insertion of $S_{X}$ removes the dependence of $T$ on $\chi$ and $S$, so that $\tilde{T}=\tilde{T}(\tilde{\pi}_{\sigma},\tilde{\pi}_{\theta},\tilde{\pi}^{i}{}_{i},\sigma,X)$.

The two constraints $\bar{\mathcal{C}}_{\tau}$ ($\tau=1,2$) in the precursor theory are linear combinations of $\mathcal{C}_i$ ($i=1,2,3$) and thus the number of new constraints to the theory is two.

With the introduction of the two supplementary constraints\footnote{The newly introduced constraints, just as the $\mathcal{C}_\tau$ and the Poisson bracket $\{\tilde{\mathcal{R}}_i(x), \mathcal{R}_0 \}$, do commute with constraints $P_X$, $P_{\chi}$, $P_S$, $S_X$, $S_{\chi}$, and $S_S$. If $\{\tilde{\mathcal{R}}_i(x), \mathcal{R}_0 \}$ did not commute with the previous set of constraint, then the $n_\tau^i$ defined by $\tilde{\mathcal{C}}_\tau = n_\tau^i \mathcal{C}_i$, being perpendicular vectors to $\{\tilde{\mathcal{R}}_i(x), \mathcal{R}_0 \}$ would then not commute with the constraints, thus not letting them commute with the newly defined constraints.} contained in $\mathcal{C}_0$ and $\mathcal{C}_i$, we have obtained enough constraints to reduce the number of physical degrees of freedom to three (or less), as intended. By the analysis of cosmological perturbation in the next sections, the model will be shown to propagate three (or more) degrees of freedom. This is enough to prove that the model has three degrees of freedom at fully nonlinear level. Alternatively, it would be possible to check whether or not new constraints are generated by inspecting the closedness of the constraint algebra, as in Eq.~(\ref{eq:closedness}). This alternative option leads, however, to more tedious calculations.

Summarizing the analysis so far, we can write more explicitly the total Hamiltonian density for the minimal theory as
\begin{eqnarray}
\mathcal{H} & = & N\sqrt{\gamma}\left[\frac{2\tilde{\pi}_{ij}\tilde{\pi}^{ij}-\tilde{\pi}^{2}}{\Mpl^{2}}-(\theta^{;i}\sigma_{;i}+P+\tilde{\pi}_{\theta}\tilde{\pi}_{\sigma})+\frac{1}{2}\chi\left(\tilde{\pi}_{\theta}^{2}-\sigma_{;i}\sigma^{;i}-2X\right)+S\,(G-\theta)-\frac{1}{2}\Mpl^{2}\,{}^{(3)\!}R\right]\nonumber \\
 &  & {}+\sqrt{\gamma}N^{i}\,(\tilde{\pi}_{\theta}\theta_{;i}+\tilde{\pi}_{\chi}\chi_{;i}+\tilde{\pi}_{\sigma}\sigma_{;i}+\tilde{\pi}_{S}S_{;i}+\tilde{\pi}_{X}\,X_{;i}-2\tilde{\pi}_{i}{}^{j}{}_{;j})+\sqrt{\gamma}\,(\bar{\lambda}_{X}\tilde{\pi}_{X}+\bar{\lambda}_{S}\tilde{\pi}_{S}+\bar{\lambda}_{\chi}\tilde{\pi}_{\chi})\nonumber \\
 &  & {}+\frac{1}{2}\sqrt{\gamma}\lambda_{\chi}\left(\sigma_{;i}\sigma^{;i}-\tilde{\pi}_{\theta}^{2}+2X\right)+\sqrt{\gamma}\lambda_{X}\left(\chi-SG_{,X}+P_{,X}\right)+\sqrt{\gamma}\lambda_{S}(\theta-G)\nonumber \\
 &  & {}-\lambda_{T}\sqrt{\gamma}\left[\frac{2G_{,X}(\tilde{\pi}_{ij}\sigma^{;i}\sigma^{;j}+X\tilde{\pi})}{\Mpl^{2}}+G_{,X}\sigma_{;i}\nabla^{i}\tilde{\pi}_{\theta}-\tilde{\pi}_{\theta}G_{,X}\sigma^{;i}{}_{;i}+\tilde{\pi}_{\theta}P_{,X}+\tilde{\pi}_{\sigma}\right]\nonumber \\
 &  & {}+\frac{m^{2}\Mpl^{2}}{2}\left[\left(\frac{\partial\mathcal{H}_{1}}{\partial\sigma}\lambda^{i}\sigma_{;i}+\frac{1}{2}\sqrt{\gamma}\Theta^{ik}\gamma_{kj}\lambda^{j}{}_{;i}+\mathcal{H}_{1}\right)+\lambda\left(\frac{\sqrt{\gamma}(2\Theta^{ij}\tilde{\pi}_{ij}-\Theta\tilde{\pi})}{2\Mpl^{2}}-\frac{\partial\mathcal{H}_{1}}{\partial\sigma}\tilde{\pi}_{\theta}\right)+N\mathcal{H}_{0}\right],
\end{eqnarray}
where we have replaced and will replace, from here on, $M = 1$ and $\tilde{\gamma}_{ij} = \delta_{ij}$.

\subsection{Lagrangian of the minimal theory}

Having the Hamiltonian, we need to perform a Legendre transformation in order to find the Lagrangian of the theory. For this purpose, on using the Hamilton equations, we find the relations between the time-derivatives of the variables and the canonical momenta, as follows
\begin{eqnarray}
\dot{\gamma}_{ij} & = & \frac{N}{\Mpl^{2}}\,(4\tilde{\pi}_{ij}-2\gamma_{ij}\tilde{\pi})+D_{j}N_{i}+D_{i}N_{j}-\frac{2\lambda_{T}G_{,X}}{\Mpl^{2}}\,(X\gamma_{ij}+\sigma_{;i}\,\sigma_{;j}) +\frac{m^2\lambda}4\,(2\Theta_{ij}-\Theta\gamma_{ij})\,,\label{eqn:gammadot}\\
\dot{\sigma} & = & -\lambda_{T}-N\,\tilde{\pi}_{\theta}+N^{i}\sigma_{;i}\,,\\
\dot{\theta} & = & N\,(\chi\tilde{\pi}_{\theta}-\tilde{\pi}_{\sigma})+N^{i}\theta_{;i}+G_{,X}\,\sigma_{;i}\lambda_{T}^{;i}-\lambda_{\chi}\tilde{\pi}_{\theta}+\lambda_{T}\!\left[\chi+G_{,XX}\sigma_{;i}\,X^{;i}-G_{,X}(S-2\sigma^{;j}{}_{;j})\right]\nonumber \\
 &  & {}-\frac{\lambda m^{2}\Mpl^{2}}{2\sqrt{\gamma}}\,\frac{\partial\mathcal{H}_{1}}{\partial\sigma}\,. \label{eqn:thetadot}
\end{eqnarray}
We now need to invert the relations (\ref{eqn:gammadot})-(\ref{eqn:thetadot}) in order to write down the (tilded) canonical momenta in terms of the first derivatives of the fields. This step leads to
\begin{eqnarray}
\tilde{\pi}_{\theta} & = & -\partial_\perp\sigma-\frac{\lambda_{T}}{N}\,,\\
\tilde{\pi}^{ij} & = & \frac{\Mpl^{2}}{2}\,(K^{ij}-K\,\gamma^{ij})-\frac{m^{2}\Mpl^{2}}{8}\,\frac{\lambda}{N}\,\Theta^{ij}+\frac{\lambda_{T}}{2N}\,G_{,X}\,[\sigma^{;i}\sigma^{;j}-\gamma^{ij}(2X+\sigma_{;k}\sigma^{;k})]\,,\\
\tilde{\pi}_{\sigma} & = &\left(\frac{\lambda_\chi}{N}-\chi\right)\partial_\perp\sigma-\partial_\perp\theta+\frac{G_{,X}}{N}\,\sigma_{;i}\lambda_{T}^{;i}+\frac{\lambda_{T}}{N}\left[\frac{\lambda_{\chi}}{N}-\chi+2G_{,X}\sigma^{;i}{}_{;i}-P_{,X}+G_{,XX}\sigma_{;i}X^{;i}\right]-\frac{\lambda m^{2}\Mpl^{2}}{2N\sqrt{\gamma}}\,\frac{\partial\mathcal{H}_{1}}{\partial\sigma}\,.
\end{eqnarray}

On writing down the Lagrangian, we find it useful to redefine appropriately the following two Lagrange multipliers as
\begin{eqnarray*}
\lambda_{\chi} & \to & N\,(\lambda_{\chi}+\chi)\,,\\
\lambda_{S} & \to & N\,(\lambda_{S}+S)\,.
\end{eqnarray*}
Furthermore, we impose, at the level of the Lagrangian, the two constraints:
$\chi=S\,G_{,X}-P_{,X}$ and $\theta=G(X)$. In other words we integrate
out the fields $\chi$ and $\theta$. We finally obtain the Lagrangian
of the Horndeski extension of the minimal theory of quasidilaton massive gravity:
\begin{eqnarray}
\mathcal{L} & = & N\sqrt{\gamma}\left\{ \frac{\Mpl^{2}}{2}\,\bigl[{}^{(3)}R+K_{ij}K^{ij}-K^{2}\bigr]+P+G_{,X}g^{\mu\nu}\partial_{\mu}X\,\partial_{\nu}\sigma\right\} \nonumber \\
 &  & {}+\lambda_{\chi}N\sqrt{\gamma}\left[\frac{\lambda_{T}}{N}\left(\partial_\perp\sigma+\frac{\lambda_{T}}{N}\right)-\frac{\lambda_{T}^{2}}{2N^2}-\frac{1}{2}\,(2X+g^{\mu\nu}\partial_{\mu}\sigma\partial_{\nu}\sigma)\right]\nonumber \\
 &  & {}+\sqrt{\gamma}\,G_{,X}\,\lambda_{T}^{;i}\sigma_{;i}\left(\partial_\perp\sigma+\frac{\lambda_{T}}{N}\right)-\frac{m^{2}\Mpl^{2}}{2}\left[\mathcal{H}_{1}+N\mathcal{H}_{0}+\frac{\partial\mathcal{H}_{1}}{\partial\sigma}\,\sigma_{;i}\lambda^{i}+\frac{1}{2}\sqrt{\gamma}\,\Theta^{jk}\gamma_{ki}\lambda^{i}{}_{;j}\right]\nonumber \\
 &  & {}+\frac{m^{4}\Mpl^{2}\lambda^{2}\sqrt{\gamma}}{64N}\,(2\Theta_{ij}\Theta^{ij}-\Theta^{2})-\frac{m^{2}\Mpl^{2}\lambda}{4}\left[2\left(\partial_\perp\sigma+\frac{\lambda_{T}}{N}\right)\frac{\partial\mathcal{H}_{1}}{\partial\sigma}+\sqrt{\gamma}K_{ij}\Theta^{ij}\right]\nonumber \\
 &  & {}-\frac{\lambda\lambda_{T}m^{2}}{4N}\,\sqrt{\gamma}\,G_{,X}\,(X\,\Theta+\Theta^{ij}\sigma_{;i}\sigma_{;j})+\lambda_{T}\sqrt{\gamma}\left\{ G_{,X}(K_{ij}\sigma^{;i}\sigma^{;j}-K\sigma_{;i}\sigma^{;i}-2X\,K)\right.\nonumber \\
 &  & {}+\left(\partial_\perp\sigma+\frac{\lambda_{T}}{N}\right)(G_{,XX}X^{;i}\sigma_{;i}+2G_{,X}\sigma^{;i}{}_{;i}-P_{,X})-G_{,X}\partial_\perp X\nonumber \\
 &  & {}-\left.\frac{1}{\Mpl^2}\frac{\lambda_{T}}{N}\,X\,G_{,X}^{2}(2\sigma_{;i}\sigma^{;i}+3X)\right\} .\label{eq:Lagr}
\end{eqnarray}
This Lagrangian (as well as the Hamiltonian) looks complicated, but it is built so that the theory possesses only three degrees of freedom (instead of six). It should be noted that $X$ here represents a general scalar field. Its definition in terms of the dynamical fields will be set by the other Lagrange multipliers of the theory, and it is not necessarily equal to $-\tfrac{1}{2}\,g^{\mu\nu}\partial_{\mu}\sigma\partial_{\nu}\sigma$, in general.

\section{The $G(X)=$const limit} \label{sec:pcase}

Although we introduced the cubic Horndeski term with the coefficient function $G(X)$ to render the minimal theory of quasidilaton massive gravity to be phenomenologically viable, in this section, as a consistency check, we shall confirm that the $G(X)={\rm const.}$ limit is regular and well-defined.

\subsection{The $P(X)=\omega\,X$ case}

In order to check whether or not the expression obtained in Eq.\ (\ref{eq:Lagr}) is consistent with the Lagrangian of the minimal quasidilaton introduced in \cite{DeFelice:2017wel}, let us first consider the case with $P=\omega\,X$ and $G(X)={\rm constant}$, where $\omega$ is a non-vanishing constant. In this case
we can directly set $G_{,X}=0=G_{,XX}$ in the Lagrangian.
\begin{itemize}
\item $X$ becomes a Lagrange multiplier, and its equation of motion gives
a constraint for $\lambda_{\chi}$, as follows
\begin{equation}
\lambda_{\chi}=\omega\,.
\end{equation}
\item On substituting this relation for $\lambda_{\chi}$ in the Lagrangian, the equation of motion for $\lambda_{T}$ leads to the
following constraint
\begin{equation}
\lambda_{T}=-\frac{m^{2}\Mpl^{2}}{2\omega\,\sqrt{\gamma}}\,\frac{\partial\mathcal{H}_{1}}{\partial\sigma}\,\lambda\,.
\end{equation}
This in particular implies that $\lambda_T$ vanishes when $\lambda$ vanishes. 
 \item After substituting this solution to the Lagrangian, we finally recover the
       theory of the minimal quasidilaton introduced in \cite{DeFelice:2017wel}.
\end{itemize}

\subsection{General $P(X)$ case}

In this case let us substitute $G_{,X}=0$, but leave $P$ as a general function of the $X$ field. In this case the equation of motion for $X$ leads to the following constraint for the Lagrange multiplier $\lambda_\chi$:
\begin{equation}
\lambda_{\chi}=P_{,X}-\frac{\lambda_{T}\,P_{,XX}}{N}\left(\partial_\perp\sigma+\frac{\lambda_{T}}{N}\right).\label{eq:lambchiHP}
\end{equation}
On the other hand, the equation of motion for the Lagrange multiplier $\lambda_{\chi}$ itself gives another constraint, which can be written as
\begin{equation}
 \mathfrak{X} = X - \frac{\lambda_{T}}{2N}\left(\frac{\lambda_{T}}{N}+2\partial_\perp\sigma\right)\,.\label{eq:sollchi}
\end{equation}
This last equation can be formally solved for $\mathfrak{X}$ in terms of the other fields.

Now we use the constraint Eq.\ (\ref{eq:lambchiHP}) into the Lagrangian to integrate out the field $\lambda_\chi$, and we find the equation of motion for the Lagrange multiplier $\lambda_T$, that we call as $\mathcal{E}_T=0$. We find that this equation is cubic in $\lambda_T$. Furthermore it explicitly contains two terms in $\mathfrak{X}$, which can be removed on using the constraint (\ref{eq:sollchi}). Now the new equation $\mathcal{E}_T=0$ has a term in $\dot\sigma^2$, which can be replaced in terms of $\mathfrak{X}$, by using the defining expression in the ADM decomposition for the latter. If then we further substitute $\mathfrak{X}$ once more in terms of the constraint (\ref{eq:sollchi}), we finally find that $\mathcal{E}_T=0$ reduces to: 
\begin{equation}
\lambda_{T}=-\frac{m^{2}\Mpl^{2}}{2\sqrt{\gamma}\,[P_{,XX}\,(2X+\sigma_{;i}\sigma^{;i})+P_{,X}]}\,\frac{\partial\mathcal{H}_{1}}{\partial\sigma}\,\lambda\,.
\end{equation}

Therefore, once again, if $\lambda=0$ then $\lambda_{T}$ also vanishes.

\section{Cosmological background} \label{sec:background}

Let us consider the equations of motion for a (flat) FLRW background for a general $P(X)$ and a general $G(X)$. In this case we can find the two modified Einstein equations, $E_{1,2}$, and the equations of motion for the fields $X,\sigma,\lambda_{\chi},\lambda_{T},\lambda$. 

\subsection{Existence of a solution with $\lambda=\lambda_T=0$}

The equation of motion for $X$, that we call $E_{X}=0$, can be solved
for $\lambda_{\chi}$. We can also solve the equation of motion for
$\lambda_{\chi}$ for $X$, and we find
\begin{equation}
X=\frac{\lambda_{T}^{2}+2\dot{\sigma}\lambda_{T}+\dot{\sigma}^{2}}{2N^{2}}\,.
\end{equation}
If we now consider the equation of motion for $\lambda_{T}$, $E_{T}=0$,
on substituting the value of $\lambda_{\chi}$ and $X$ obtained above,
we find it can be rewritten as
\begin{equation}
E_{T}=\lambda\sum_{j=0}^{4}f_{j}(N,\dot{\sigma},\sigma,a,\lambda_{T})(\lambda_{T})^{j}+\sum_{i=1}^{7}g_{i}(N,\dot{\sigma},\dot{a},a,\lambda_{T})(\lambda_{T})^{i}=0\,.\label{eq:lTeq}
\end{equation}
This shows that the relation between $\lambda$ and $\lambda_{T}$
is an algebraic one. Later on, we will use this equation so that we
can consider $\lambda$ as $\lambda=\lambda(\lambda_{T},\dots)$ (without
substituting its expression directly into the other equations of motion),
where the other variables do not contain time derivative of $\lambda$
or $\lambda_{T}$. Notice that $\lim_{\lambda_{T}\to0}\lambda=0$,
in general. We have so far used two equations of motion to define
$\lambda_{\chi}$, $X$ in terms of other fields, and found a relation
between $\lambda$ and $\lambda_{\chi}$ using the equation of motion
$E_{T}$.

Let us now reset all the definitions of $X$ and $\lambda_{\chi}$,
and build an expression by using also the other equations of motion
as in

\begin{eqnarray*}
E_{0} & \equiv & \dot{E}_{1}+\frac{3\dot{a}}{a}\,(E_{1}-E_{2})+\beta_{3}\dot{E}_{\lambda}-\frac{m^{2}\Mpl}{2}\,e^{(1+\alpha)\,\frac{\sigma}{\Mpl}}\,E_{\lambda}+\beta_{5}E_{T}+\beta_{6}\dot{E}_{T}+\beta_{7}E_{\sigma}+\beta_{8}E_{\chi}+\beta_{9}\dot{E}_{\chi}+\beta_{10}E_{X}=0\,,
\end{eqnarray*}
where the $\beta$'s are (finite) functions to be determined. We will
choose $\beta_{5}=0$, and look for $\beta_{k}$ ($k=3,6,9,10)$,
so that the terms in $\dot{\lambda}$, $\dot{\lambda}_{T}$, $\dot{\lambda}_{\chi}$,
$\lambda_{\chi}$ all vanish. Furthermore we set other two conditions
on $\beta_{7,8}$ so as to make vanish the term which is found on
calculating $\lim_{\lambda,\lambda_{T}\to0}E_{0}$. It is possible
indeed to build such an expression, which can be written as
\begin{equation}
E_{0}=h^{(2)}\lambda^{2}+\lambda\sum_{j=0}^{5}h_{j}^{(1)}(\lambda_{T})^{j}+\sum_{i=1}^{7}h_{i}^{(0)}(\lambda_{T})^{i}=0\,.
\end{equation}
On combining this equation with Eq.\ (\ref{eq:lTeq}) shows that indeed there exists a solution with 
\begin{equation}
 \lambda = \lambda_{T}=0\,. \label{eqn:lambdaT=0}
\end{equation}

\subsection{Uniqueness of the solution with $\lambda=\lambda_T=0$}

In this subsection we show that the solution (\ref{eqn:lambdaT=0}) not only exists but is also unique, based on a mini-superspace analysis.

Let us consider the Lagrangian density, $\mathcal{L}$, in Eq.\ (\ref{eq:Lagr}) and evaluate it in the mini-superspace (i.e.\ $\mathcal{L}\to\mathfrak{L}$), for which the variables
\begin{eqnarray}
\gamma_{ij} & = & a(t)^{2}\,\delta_{ij}\,,\\
\sigma & = & \sigma(t)\,,\\
X & = & X(t)\,,
\end{eqnarray}
are the only ones not to be Lagrange multipliers. Then, for these three variables ($a,\sigma,X$),
we are able to find and invert the relation between their time derivatives and the canonical momenta. 

After performing the field redefinition, $\lambda_{\chi}\to\lambda_{\chi}/N$,
we find the following result for the mini-superspace Hamiltonian,
$\mathfrak{H}=\pi_{a}\dot{a}+\pi_{X}\dot{X}+\pi_{\sigma}\dot{\sigma}-\mathfrak{L}$:
\begin{eqnarray}
\mathfrak{H} & = & \left(c_{{3}}+\frac{3c_{{2}}}{a}{\rm e}^{\frac{\sigma}{\Mpl}}+\frac{3c_{{1}}}{a^{2}}{\rm e}^{\frac{2\sigma}{\Mpl}}+\frac{c_{{0}}}{a^{3}}{\rm e}^{\frac{3\sigma}{\Mpl}}\right)a^{3}{\rm e}^{\frac{\left(1+\alpha\right)\sigma}{\Mpl}}\,\frac{m^{2}\Mpl^{2}}{2}\nonumber \\
 &  & {}-{\rm e}^{{\frac{\left(1+\alpha\right)\sigma}{\Mpl}}}\left[\left(c_{{3}}+\frac{2}{a}{\rm e}^{\frac{\sigma}{\Mpl}}c_{{2}}+\frac{c_{{1}}}{a^{2}}{\rm e}^{\frac{2\sigma}{\Mpl}}\right)\frac{a\,m^{2}}{4}\pi_{a}\right.\nonumber \\
 &  & {}+\left.\left(c_{{3}}\left(1+\alpha\right)+\frac{3c_{{2}}\left(2+\alpha\right)}{a}{\rm e}^{\frac{\sigma}{\Mpl}}+\frac{3c_{{1}}\left(3+\alpha\right)}{a^{2}}{\rm e}^{\frac{2\sigma}{\Mpl}}+\frac{c_{{0}}\left(4+\alpha\right)}{a^{3}}{\rm e}^{\frac{3\sigma}{\Mpl}}\right)\frac{\Mpl m^{2}}{2G_{,X}}\pi_{X}\right]\lambda\nonumber \\
 &  & {}+\left[\left(\frac{c_{{1}}}{a^{3}}{\rm e}^{\frac{3\sigma}{\Mpl}}+\frac{3}{a^{2}}\,{\rm e}^{\frac{2\sigma}{\Mpl}}c_{{2}}+\frac{3}{a}{\rm e}^{\frac{\sigma}{\Mpl}}c_{{3}}+c_{{4}}\right)\frac{a^{3}m^{2}\Mpl^{2}}{2}-\frac{1}{12}\,\frac{\pi_{{a}}^{2}}{\Mpl^{2}a}-\frac{\pi_{{\sigma}}\pi_{{X}}}{G_{,X}a^{3}}-a^{3}P\right]N\nonumber \\
 &  & {}+\left(a^{3}X-\frac{\pi_{X}^{2}}{2G_{,X}^{2}a^{3}}\right)\lambda_{\chi}-\left(\frac{G_{,X}Xa\pi_{a}}{\Mpl^{2}}+\frac{P_{,X}\pi_{X}}{G_{,X}}+\pi_{\sigma}\right)\lambda_{T}\,.\label{eq:Hmini}
\end{eqnarray}
If we call by $\mathfrak{R}_{0}$ and $\mathfrak{C}_{\chi}$, respectively the two constraints
set by the Lagrange multipliers $N$ and $\lambda_{\chi}$ in the
Hamiltonian, then we find that
\begin{eqnarray}
\dot{\mathfrak{R}}_{0} & = & \{\mathfrak{R}_{0},\mathfrak{H}\}\approx\zeta_{1}\lambda+\zeta_{2}\lambda_{T}=0\,,\\
\dot{\mathfrak{C}}_{\chi} & = & \{\mathfrak{C}_{\chi},\mathfrak{H}\}\approx\zeta_{3}\lambda+\zeta_{4}\lambda_{T}=0\,,
\end{eqnarray}
then \textendash{} unless some extra dynamical constraint (the vanishing
of the determinant $\zeta_{1}\zeta_{4}-\zeta_{2}\zeta_{3}$ besides
the known constraints), which will drastically reduce the number of
available backgrounds, is imposed \textendash{} we find the following
unique solution
\begin{equation}
\lambda\approx0\approx\lambda_{T}\,.
\end{equation}
This argument proves the uniqueness of such a solution for the FLRW background.
The time-derivatives of the other two constraints, $\mathfrak{C}_{\lambda}$,
and $\mathfrak{C}_{T}$ (set by $\lambda$ and $\lambda_{T}$, respectively),
can be used, in principle, to determine the values of $N$, and $\lambda_{\chi}$
in terms of the other dynamical variables. 

\subsection{Background equations of motion with $\lambda=\lambda_{T}=0$}

In the following we will study the only allowed solution for a general
FLRW background, namely $\lambda=0=\lambda_{T}$. In this case we
find that
\begin{eqnarray}
X & = & \frac{\dot{\sigma}^{2}}{2N^{2}}=\frac{1}{2}\,\Sigma^2\,,\\
\lambda_{\chi} & = & P_{,X}+G_{,X}\left(\frac{\dot{\Sigma}}{N}+3\,H\,\Sigma\right),
\end{eqnarray}
where we have defined
\begin{eqnarray}
\Sigma & \equiv & \dot{\sigma}/N=\Mpl\left(\frac{\dot{\mathcal{X}}}{N\mathcal{X}}+H\right)\,,\\
H & \equiv & \frac{\dot{a}}{Na}\,,\\
\mathcal{X} & \equiv & \frac{{\rm e}^{\sigma/\Mpl}}{a}\,,\\
r & \equiv & \frac{a\,{\rm e}^{\alpha\sigma/\Mpl}}{N}\,.
\end{eqnarray}
In this case we find the following vacuum equations of motion
\begin{eqnarray}
E_{1} & = & 3\,H^{2}\Mpl^{2}+P-P_{,X}\Sigma^{2}-3HG_{,X}\Sigma^{3}-\frac{1}{2}\left(c_{4}+3c_{3}\mathcal{X}+3c_{2}\mathcal{X}^{2}+c_{1}\mathcal{X}^{3}\right)\Mpl^{2}m^{2}=0\,,\label{eq:E1}\\
E_{2} & = & \left(\frac{2\dot{H}}{N}+3H^{2}\right)\Mpl^{2}+P-\frac{\Sigma^{2}G_{,X}\dot{\Sigma}}{N}-\frac{1}{2}\left[c_{4}+2c_{3}\mathcal{X}+c_{2}\mathcal{X}^{2}+r\mathcal{X}\left(c_{3}+2c_{2}\mathcal{X}+c_{1}\mathcal{X}^{2}\right)\right]\Mpl^{2}m^{2}=0\,,\label{eq:E2}\\
E_{\sigma} & = & \frac{1}{2}\left\{ \left[c_{0}\left(4+\alpha\right)\mathcal{X}^{3}+3c_{1}\left(3+\alpha\right)\mathcal{X}^{2}+3c_{{2}}\left(2+\alpha\right)\mathcal{X}+c_{{3}}\left(1+\alpha\right)\right]r+3(c_{{3}}+2c_{{2}}\mathcal{X}+c_{{1}}\mathcal{X}^{2})\right\} \mathcal{X}\Mpl\,m^{2}\nonumber \\
 &  & {}+3\,H\Sigma\,(P_{,X}+3\,HG_{,X}\Sigma)+3\,\Sigma^{2}G_{,X}\,\frac{\dot{H}}{N}+(3\,HG_{,{\it XX}}\Sigma^{3}+6\,H\Sigma\,G_{,X}+P_{{\it ,XX}}\Sigma^{2}+P_{,X})\,\frac{\dot{\Sigma}}{N}=0\,,\\
E_{\lambda} & = & 3\left(\mathcal{X}^{2}c_{1}+2\,\mathcal{X}\,c_{2}+c_{3}\right)H\Mpl+\left[c_{0}\mathcal{X}^{3}\left(4+\alpha\right)+3\mathcal{X}^{2}c_{1}\left(3+\alpha\right)+3\mathcal{X}\,c_{2}\left(2+\alpha\right)+c_{3}\left(1+\alpha\right)\right]\Sigma=0\,.
\end{eqnarray}

It should be noticed that the equation of motion $E_{\lambda}=0$ can
also be written as
\begin{equation}
\frac{d}{dt}\left[a^{4+\alpha}\mathcal{X}^{1+\alpha}\,J(\mathcal{X})\right]=0\,,\label{eq:eqlX}
\end{equation}
where
\begin{equation}
J\equiv c_{0}\mathcal{X}^{3}+3c_{1}\mathcal{X}^{2}+3c_{2}\mathcal{X}+c_{3}\,.\label{eq:Jeq}
\end{equation}
Therefore this theory always admits an attractor, which if we neglect
the strongly-coupled case $\mathcal{X}=0$, reduces, if $\alpha\neq-4$,
to setting the constraint
\begin{equation}
J(\mathcal{X})=0\,.
\end{equation}
On the other hand, if $\alpha=-4$, then the solution of Eq.\ (\ref{eq:eqlX})
reduces to $\mathcal{X}$ being (any positive) constant.

On combining the first and the second Einstein equations, we also
find that 
\begin{equation}
r-1=\frac{2\,(P_{,X}+3\,H^{2}\Mpl\,G_{,X})}{\mathcal{X}\,(c_{1}\mathcal{X}^{2}+2c_{2}\mathcal{X}+c_{3})}\,\frac{H^{2}}{m^{2}}\,,
\end{equation}
so that $r\to1$ corresponds, in general, to a Minkowski limit (and
vice-versa).

Finally, we impose on the Friedmann equation that at constant $\mathcal{X}$, an increase of the bare cosmological constant $\Lambda$ (which here is proportional to $m^2\,c_4$ and the possibly existent constant term $P(X=0)$) always leads to an increase of the Hubble parameter $H$. This implies that $(\partial\Lambda/\partial H)_{\mathcal{X}}>0$, which, in turn, leads to
\begin{equation}
  6-P_{,X}-3\Mpl^{3}\,H^{4}G_{,{\it XX}}-\Mpl^{2}H^{2}P_{,{\it XX}}-12\Mpl\,H^{2}\,G_{,X}>0\,.
 \end{equation}

\section{Linear Perturbation} \label{sec:stability}

In the following, we will study the dynamics of the perturbation to
a de Sitter background, in order to address the issue of the stability
of the theory. Since here we have defined the theory, we will only
address the de Sitter background, which we impose to correspond to
the final state of the universe evolution. Looking at the equations
of motion, $X={\rm const}.$ leads to $\Sigma=\Mpl\,H={\rm const}$,
which is consistent with a de Sitter background. In the following
we will only discuss scalar and tensor perturbations, as the gravity
vector modes are absent in this model, due to the constraint imposed
by the $\lambda^{i}$ field.

\subsection{Scalar perturbations}

We will write the lapse, the shift, and the three-dimensional metric as follows 
\begin{eqnarray}
N & = & N(t)\,(1+\delta N)\,,\\
N_{i} & = & N(t)\,\partial_{i}\psi\,,\\
ds^{2} & = & a(t)^{2}\,[(1+2\Phi)\delta_{ij}+2\partial_{i}\partial_{j}s]\,dx^{i}\,dx^{j}\,.
\end{eqnarray}
Furthermore we perturb the scalar fields as in
\begin{eqnarray}
\sigma & = & \sigma(t)+\delta\sigma\,,\\
X & = & \frac{1}{2}\,\Mpl^{2}H^{2}+\delta X\,,\\
\lambda & = & \delta\lambda\,,\\
\lambda_{T} & = & \delta\lambda_{T}\,,\\
\lambda_{\chi} & = & P_{,X}+3\Mpl H^{2}G_{,X}+\delta\lambda_{\chi}\,,
\end{eqnarray}
and the vector Lagrange multiplier $\lambda^{i}$ as
\begin{equation}
\lambda^{i}=\frac{\delta^{ik}}{a^{2}}\,\partial_{k}\delta\lambda_{V}\,.
\end{equation}

\subsubsection{Case $\alpha\protect\neq-4$}

Here we study the case $J=0$. In this case we can set $c_{4}=0$ and consider $P(X=0)=0$
(removing a bare cosmological constant, looking for a self-accelerating
solution), and solve the equations of motion $(\ref{eq:E1})$, ($\ref{eq:E2}$),
and ($\ref{eq:Jeq}$) for, say, $c_{0},c_{1}$ and $c_{2}$. In this
case, we find that all the background equations of motion are satisfied.

The analysis of the perturbation can be done, in Fourier space, by
performing the following steps.
\begin{enumerate}
\item We use the constraint imposed by $\delta\lambda_{V}$ to write $\Phi$
in terms of $\delta\sigma$.
\item We can use the equation of motion for $\delta\lambda_{\chi}$ to write
$\delta\lambda_{T}$ in terms of $\delta N,\delta X$, and $\dot{\delta\sigma}$.
\item We can use the equation of motion for $\psi$ to write $\delta N$
in terms of $\dot{\delta\sigma},\delta\sigma,$ and $\delta\lambda$.
\item We can use the equation of motion for $\delta\lambda$ to write $\delta X$
in terms of $s$ and $\delta\sigma$.
\end{enumerate}
After integrating out the field $\delta X$, one is left with a Lagrangian
which, after a few integrations by parts, can be written as
\begin{equation}
\mathcal{L}=N\,a^{3}\left[\mathcal{S}_{1}\,s_{\bm{k}}s_{-\bm{k}}+\mathcal{S}_{2}\,s_{\bm{-k}}\,\frac{\dot{\delta\sigma}_{\bm{k}}}{N}+\mathcal{S}_{3}\,s_{\bm{-k}}\,\delta\sigma_{\bm{k}}+\mathcal{S}_{4}\,(\delta\sigma_{\bm{k}})(\delta\sigma_{-\bm{k}})\right].
\end{equation}
It is clear that after integrating out $s$ (or $\delta\sigma$),
one will have only one single propagating scalar field, as required
by the construction of the theory.

After integrating out the field $s$, we study the high-$k$ behavior
of the previously written Lagrangian. We find that
\begin{eqnarray}
\mathcal{S}_{1} & = & \mathcal{S}_{1}^{(4)}k^{4}\,,\\
\mathcal{S}_{2} & = & \mathcal{S}_{2}^{(2)}k^{2}\,,\\
\mathcal{S}_{3} & = & \mathcal{S}_{3}^{(4)}\frac{k^{4}}{a^{2}}+\mathcal{S}_{3}^{(2)}k^{2}\,,\\
\mathcal{S}_{4} & = & \mathcal{S}_{4}^{(4)}\frac{k^{4}}{a^{4}}+\mathcal{S}_{4}^{(2)}\frac{k^{2}}{a^{2}}+\mathcal{O}(k^0)\,,
\end{eqnarray}
where the coefficients of this expansion, $\mathcal{S}_{i}^{(2,4)}$
($i=1,\dots,4$), are all constants on the background, and $k$-independent.
Therefore, in the subhorizon approximation, the Lagrangian reduces
to
\begin{equation}
\mathcal{L}\approx N\,a^{3}\left[-\frac{[\mathcal{S}_{2}^{(2)}]^{2}}{4\mathcal{S}_{1}^{(4)}}\,\frac{(\dot{\delta\sigma}_{\bm{k}})(\dot{\delta\sigma}_{-\bm{k}})}{N^{2}}+\left(\mathcal{S}_{4}^{(2)}-\frac{\mathcal{S}_{3}^{(4)}\mathcal{S}_{3}^{(2)}}{2\mathcal{S}_{1}^{(4)}}+\frac{H\mathcal{S}_{2}^{(2)}\mathcal{S}_{3}^{(4)}}{4\mathcal{S}_{1}^{(4)}}\right)\frac{k^{2}}{a^{2}}\,(\delta\sigma_{\bm{k}})(\delta\sigma_{-\bm{k}})\right]\,. \label{eqn:Ldeltasigma}
\end{equation}
where we have used the fact that 
\begin{equation}
\mathcal{S}_{4}^{(4)}-\frac{[\mathcal{S}_{3}^{(4)}]^{2}}{4\mathcal{S}_{1}^{(4)}}=0\,. \label{eqn:k4coef} 
\end{equation}
The relation (\ref{eqn:k4coef}) ensures that terms proportional to $k^4$ cancel each other in (\ref{eqn:Ldeltasigma}) and that the perturbations have the standard high-$k$ propagation structure, i.e.\ the linear dispersion relation. We can then rewrite the subhorizon Lagrangian, in position space, as
\begin{equation}
\mathcal{L}\approx Na^{3}\,Q\left[\frac{(\dot{\delta\sigma})^{2}}{N^{2}}-\,\frac{c_{s}^{2}}{a^{2}}\,(\partial_{i}\delta\sigma)^{2}\right],
\end{equation}
where the no-ghost condition can be written as
\begin{equation}
Q=-\frac{[\mathcal{S}_{2}^{(2)}]^{2}}{4\mathcal{S}_{1}^{(4)}}>0\,.\label{eq:QsJ}
\end{equation}
This relation is equivalent to the following one
\begin{equation}
P_{,X}+\Mpl^{2}H^{2}P_{,{\it XX}}+3\,\Mpl^{3}H^{4}G_{,{\it XX}}+\frac{3}{2}\Mpl^{2}\,H^{4}G_{,X}^{2}+6\Mpl\,H^{2}\,G_{,X}>0\,.\label{eq:NoGhSc}
\end{equation}
Furthermore, we need to impose also the condition $\lambda_{\chi}=P_{,X}+3\Mpl H^{2}G_{,X}\neq0$, as, otherwise, a strong coupling regime arises.

On the other hand, the condition of the non-negativity of the speed of propagation squared reduces
to
\begin{equation}
c_{s}^{2}=\frac{4\mathcal{S}_{1}^{(4)}}{[\mathcal{S}_{2}^{(2)}]^{2}}\left(\mathcal{S}_{4}^{(2)}-\frac{\mathcal{S}_{3}^{(4)}\mathcal{S}_{3}^{(2)}}{2\mathcal{S}_{1}^{(4)}}+\frac{H\mathcal{S}_{2}^{(2)}\mathcal{S}_{3}^{(4)}}{4\mathcal{S}_{1}^{(4)}}\right) \geq 0\,.
\end{equation}
The explicit forms for the expressions of $Q$ and $c_s^2$ can be found in appendix \ref{apx:Q-c2s}.

\subsubsection{Case $\alpha=-4$}

In this case, $\mathcal{X}={\rm constant}$ (and any positive value
is allowed), and we can further set $c_{4}=0$, and solve Eqs.\ (\ref{eq:E1})
and (\ref{eq:E2}) for, say, $c_{1}$ and $c_{2}$. The analysis simplifies
and, along the same lines of the previous case, one finds that one single scalar field propagates with a reduced Lagrangian which
can be written as
\begin{equation}
\mathcal{L}=Na^{3}\,Q\left[\frac{(\dot{\delta\sigma})^{2}}{N^{2}}-\frac{c_{s}^{2}}{a^{2}}\,(\partial_{i}\delta\sigma)^{2}-\mu_{s}^{2}\,(\delta\sigma)^{2}\right].
\end{equation}
Although the value of $Q$ -- which can be found in appendix \ref{apx:Q-c2s} -- for the $\alpha=-4$ case is different
from the value given in Eq.\ (\ref{eq:QsJ}), we find that the condition
$Q>0$ still gives the same relation written in (\ref{eq:NoGhSc}),
but, in order to avoid the strong coupling regime one now requires
\begin{equation}
3\,H^{4}\Mpl^{3}\,G_{,{\it XX}}+\Mpl^{2}H^{2}P_{,{\it XX}}+12\Mpl\,H^{2}\,G_{,X}+P_{,X}-6\neq0\,.
\end{equation}
As for the speed of propagation, one finds the following expression
\begin{equation}
c_{s}^{2}=\frac{(P_{,X}-2\Mpl\,H^{2}\,G_{,X}+10)\,(6\Mpl^{3}\,H^{4}G_{,{\it XX}}+3\Mpl^{2}\,H^{4}G_{,X}^{2}+2\Mpl^{2}\,H^{2}P_{,{\it XX}}+12\Mpl\,H^{2}\,G_{,X}+2\,P_{,X})}{2\,(3\Mpl^{3}\,H^{4}G_{,{\it XX}}+\Mpl^{2}H^{2}P_{,{\it XX}}+12\Mpl\,H^{2}\,G_{,X}+P_{,X}-6)^{2}}\geq0\,.
\end{equation}

It should be noted that, differently from the simplest minimal quasidilaton theory introduced in \cite{DeFelice:2017wel}, the scalar modes can have a non-negative mass, so that the background can be completely stable against scalar perturbations, as we have
\begin{equation}
\mu_{s}^{2}=\frac{3}{2}\,\frac{(P_{,X}+6)\,(P_{,X}+3\Mpl\,H^{2}\,G_{,X})\,H^{2}}{P_{,X}-6+\Mpl^{2}H^{2}P_{,{\it XX}}+3\Mpl^{3}\,H^{4}G_{,{\it XX}}+12\Mpl\,H^{2}\,G_{,X}}\,,
\end{equation}
which is not necessarily negative, in general.

\subsection{Tensor perturbations}

Let us perturb the three dimensional metric as in
\begin{equation}
ds^{2}=a(t)^{2}(\delta_{ij}+h_{ij})\,dx^{i}\,dx^{j}\,,
\end{equation}
where
\begin{equation}
h_{ij}=h_{+}\,\epsilon_{ij}^{+}+h_{\times}\,\epsilon_{ij}^{\times}\,,
\end{equation}
and the two polarization tensors, $\epsilon^{{+,\times}}$, are symmetric,
trace-less, and satisfying ${\rm Tr}[\epsilon^{{+}}\epsilon^{{+}}]=1={\rm Tr}[\epsilon^{{\times}}\epsilon^{{\times}}]$
and ${\rm Tr}[\epsilon^{{+}}\epsilon^{{\times}}]=0={\rm Tr}[\epsilon^{{\times}}\epsilon^{{+}}]$.
In this case we can write the Lagrangian for the perturbations as
\begin{equation}
\mathcal{L}=\frac{\Mpl^{2}}{8}\sum_{\lambda={+,\times}}Na^{3}\left[\frac{\dot{h}_{\lambda}^{2}}{N^{2}}-\,\frac{1}{a^{2}}\,(\partial_{i}h_{\lambda})^{2}-\mu_{T}^{2}\,h_{\lambda}^{2}\right],
\end{equation}
where
\begin{equation}
  \mu_{T}^{2}=\frac{c_{3}\,m^{2}\,(r-1)\,\mathcal{X}}{2}
  +\frac{m^2\,r^2\mathcal{X}\,(c_{1}\mathcal{X}^{2}+2c_{2}\mathcal{X}+c_{3})}{2}
  -\frac{(r-1)\,(P+3\,\Mpl^{2}H^{2})}{\Mpl^{2}}\,.
\end{equation}

Therefore it is possible to find a parameter space for which also the tensor modes are stable on the de Sitter background. It should be noted that in the Minkowski limit, the mass of the tensor modes does not vanish in general, as long as $m\neq0$.

\section{Summary and discussion}  \label{sec:conclusion}

In this paper we have discussed the Horndeski extension of the minimal theory of quasidilaton massive gravity, which has one scalar degree of freedom and two massive tensor modes. A Horndeski term has been introduced into the Lagrangian in order to implement the Vainshtein mechanism to the quasidilaton scalar field, so that GR can be recovered at short scales (i.e.\ solar system scales). In fact, we need to screen only the quasidilaton scalar field at solar system scales, because that is the only one scalar degree of freedom propagating in the theory. Moreover, it is well known that a cubic Horndeski term is able to implement the Vainshtein mechanism (see e.g.\ \cite{Nicolis:2008in,DeFelice:2011th,Narikawa:2013pjr}) at short scales.

On the other hand, at very large scales, the tensor modes acquire a mass which is also affected by the quasidilaton scalar field, and the universe has a self-accelerating de Sitter attractor (even in the presence of dust/radiation fluids).

We have firstly rewritten the standard cubic Horndeski Lagrangian by introducing auxiliary fields so that it does not contain second derivatives of fields. (The theory possesses three degrees of freedom. Of course this was well expected as the Horndeski Lagrangian only gives rise to second order equations of motion for the metric and the scalar. See Appendix~\ref{sec:cubichorndeski} for details of the Hamiltonian analysis.) Then we have added a Lorentz-breaking graviton mass term which describes the interaction between the quasidilaton field and the metric. This new ``precursor'' Lagrangian term is able to endow the tensor modes with a non-zero mass, and, in general, also introduces one extra-scalar degree of freedom (leftover from the three extra degrees in the standard massive gravity), so that in total one has four propagating degrees (two scalars, two tensors).  This number of degrees (four) is due to the fact that the precursor action breaks Lorentz invariance, so that we may have a number of degrees of freedom different from the value which would be imposed by Lorentz symmetry. The Lorentz breaking length-scale is of order of the cosmological horizon, so that the theory can be responsible for the acceleration of the universe, and at the same time, at short scales it should reduce to GR (apart from the quasidilaton dynamics, for which we have added a cubic Horndeski term in order to screen it).

We have pushed forward by reducing the dynamical degrees of freedom (in order to have a minimal theory and a simpler phenomenology) and we have added, in a non-linear way, two extra second-class constraints which are able to remove the extra gravity scalar degree of freedom, without killing the cosmological background solutions of the precursor theory. In this way, we are left with only three degrees of freedom (one scalar \textendash{} the quasidilaton \textendash{} and two massive gravitational waves).

Once we have defined the Hamiltonian, we have derived its Lagrangian via a Legendre transformation and applied it to study the FLRW background equations of motion. We have found the existence of a late-time de Sitter (possibly Minkowski) attractor.  Then, we have analyzed the stability of this de Sitter background, i.e.\ the final state of our universe, according to the dynamics of this theory. We have found that there are several possibilities and in general the background will be stable. Together with the fact that the cubic Horndeski term is able to endow the quasidilaton field with a successful Vainshtein mechanism at solar system scales, and the property that the gravitational waves mass does not vanish in general (even on Minkowski), we expect this theory to have an interesting phenomenology which needs to be studied in the near future.

In the present paper, we have not studied the details of the Vainshtein mechanism since, as already mentioned above, it is well known that the cubic Horndeski term is able to implement the Vainshtein mechanism at short scales. As a future work it is intriguing to confirm that the extension introduced in this paper for the minimal theory of quasidilaton massive gravity is indeed effective enough to pass solar system constraints.

\acknowledgments
ADF was supported by JSPS KAKENHI Grant Numbers 16K05348, 16H01099. The work of SM was supported by Japan Society for the Promotion of Science (JSPS) Grants-in-Aid for Scientific Research (KAKENHI) No. 17H02890, No. 17H06359, No. 17H06357, and by World Premier International Research Center Initiative (WPI), MEXT, Japan.

\appendix

\section{Equivalence at the level of equations of motion}
\label{app:eom-equivalence}

In this appendix we shall show the equivalence of the two systems (\ref{eq:lagrangian_pre_scalarpart}) and (\ref{eq:lagrangian_standard_horndeski}) at the level of equations of motion.

Considering the presence of the graviton mass term, the equation of motion of the (quasidilaton) scalar field becomes 
\begin{align}
&E_\sigma \equiv \chi\Box\sigma + \Box\theta + g^{\mu\nu}\nabla_\mu\chi\nabla_\nu\sigma - \frac{1}{N\sqrt{\gamma}}\frac{\partial\mathcal{L}_m}{\partial\sigma}= 0\,.
\end{align}
By replacing the solutions for the auxiliary fields' equations into the equation for $\sigma$ we obtain the usual equation
\begin{equation}
F_\mathfrak{,X}\Box\sigma - G_\mathfrak{,X}\Box(\mathfrak{X})- G_{,\mathfrak{X}\mathfrak{X}}g^{\mu\nu}\nabla_\mu\mathfrak{X}\nabla_\nu\mathfrak{X} + g^{\mu\nu}F_{,\mathfrak{X}\mathfrak{X}}\nabla_\mu\mathfrak{X}\nabla_\nu\sigma + \frac{1}{N\sqrt{\gamma}}\frac{\partial\mathcal{L}_m}{\partial\sigma}= 0\,,
\end{equation}
where 
\begin{equation}
G_{,\mathfrak{X}} \equiv G_{,X} \biggr\rvert_{X = \mathfrak{X}}\,,\quad G_{,\mathfrak{X} \mathfrak{X}} \equiv G_{,XX} \biggr\rvert_{X = \mathfrak{X}}\,,\quad F_{,\mathfrak{X}} \equiv F_{,X} \biggr\rvert_{X = \mathfrak{X}}\,,\quad F_{,\mathfrak{X}\mathfrak{X}} \equiv F_{,XX} \biggr\rvert_{X = \mathfrak{X}}\,.
\end{equation}
We have thus established the equivalence between the two Lagrangian densities (\ref{eq:lagrangian_pre_scalarpart}) and (\ref{eq:lagrangian_standard_horndeski}) at the level of equations of motion.

\section{Explicit Expressions for $Q$ and $c_s^2$}
\label{apx:Q-c2s}

As for the scalar perturbations, we write in the following the explicit form of the no-ghost condition and the squared speed of propagation in the case $\alpha\neq-4$:
\begin{eqnarray}
  Q &=& \frac12\,\frac{\lambda_\chi^2\,\Xi_1^2}{\Xi\,\Xi_2^2\,\Xi_3^2}\,,\\
  c_s^2 &=& \frac{\Xi\,\Xi_3\,\Xi_4}{\lambda_\chi\,\Xi_1^2}\,,\\
  \lambda_{\chi} & = & P_{,X}+3\Mpl H^{2}G_{,X}\,,\\
  \Xi &=& 3\Mpl^3H^4G_{,XX}+\tfrac32\Mpl^2H^4G_{,X}^2+P_{,XX}\Mpl^2H^2+6G_{,X}\Mpl H^2+P_{,X}\,,\\
  \Xi_1 &=& 3\,[(\alpha^2+8\alpha+40)r-\alpha^2-5\alpha+20]\Mpl^4H^6G_{,X}^2
           + 12\bigl\{\tfrac52(\tfrac1{10}\alpha+r+\tfrac75)\Mpl^2H^2P_{,X}\nonumber\\
    &&{}+\tfrac32H^6\Mpl^5(r+1)G_{,XX}+\tfrac12H^4\Mpl^4(r+1)P_{,XX}+(2-2r)P\nonumber\\
    &&{}   +\bigl[[(\alpha^2+8\alpha+7)r-19-\alpha^2-\tfrac{19}2\alpha]H^2 +m^2c_{3}\mathcal{X}(r-1)\bigr]\Mpl^2\bigr\}\Mpl H^2G_{,X}
       +2\Mpl^2H^2(r+1)P_{,X}^2\nonumber\\
    &&{} +\bigl\{6H^6\Mpl^5(r+1)G_{,XX}+(2-2r)P +2H^4\Mpl^4(r+1)P_{,XX} \nonumber\\
    &&{} +\bigl[[(2\alpha^2+16\alpha+14)r-2\alpha^2-22\alpha-62]H^2+m^2c_{3}\mathcal{X}(r-1)\bigr]\Mpl^2\bigr\}\,P_{,X}\nonumber\\
    &&{} +3(r-1)\,\bigl\{\tfrac13\Mpl^2\bigl[[(2\alpha^2+16\alpha+26)H^2+\mathcal{X}m^2c_{3}]\Mpl^2-2P\bigr]H^2(P_{,XX}+3\Mpl H^2 G_{,XX})\nonumber\\
    &&{} +4P+(12H^2-2\mathcal{X}c_{3}m^2)\Mpl^2\bigr\}\,,\\
  \Xi_2 &=& 6\Mpl^3H^4(r+1)G_{,X}+2\Mpl^2H^2(r+1)P_{,X}
            +[\Mpl^2(c_{3}m^2\,\mathcal{X}-6H^2)-2P](r-1)\,,\\
  \Xi_3 &=& \Mpl [(4+\alpha)r-\alpha-1]H^2G_{,X}+P_{,X}+2(\alpha+4)(1-r)\,,\\
  \Xi_4 &=& -36H^{10}\Mpl^7(r+1)[(4+\alpha)r^2+(-2\alpha-6)r-\alpha-2]G_{,X}^3\nonumber\\
    &&{}-12H^6\Mpl^4\bigl\{2H^2\Mpl^2(r+1)[(4+\alpha)r^2+(-2\alpha-\tfrac{15}2)r-\alpha-\tfrac72]P_{,X}\nonumber\\
    &&{} +[(-2\alpha-8)r^3+(6\alpha+20)r^2-4\alpha-12]P\nonumber\\
    &&{} +\Mpl^2\bigl[[(-12\alpha-48)r^3+(-6\alpha^2-24\alpha-42)r^2
         +(\tfrac12\alpha^3+92+12\alpha^2+78\alpha)r+\tfrac32\alpha^2+18\alpha-\tfrac12\alpha^3-2]H^2\nonumber\\
    &&{} +\mathcal{X}c_{3}[(4+\alpha)r^2+(-2\alpha-6)r-2\alpha-6]m^2(r-1)\bigr]\bigr\}\,G_{,X}^2\nonumber\\
    &&{}-H^2\Mpl \Bigl(4H^4\Mpl^4(r+1)[(4+\alpha)r^2+(-2\alpha-12)r-\alpha-8]P_{,X}^2\nonumber\\
    &&{}+4H^2\Mpl^2\bigl\{[(-2\alpha-8)r^3+(6\alpha+26)r^2-4\alpha-18]P\nonumber\\
    &&{}+\Mpl^2\bigl[[(-18\alpha-72)r^3+(-12\alpha^2-66\alpha-126)r^2
       +(\tfrac12\alpha^3+152+18\alpha^2+132\alpha)r+9\alpha^2+72\alpha-\tfrac12\alpha^3+46]H^2\nonumber\\
    &&{}+[(4+\alpha)r^2+(-2\alpha-9)r-2\alpha-9]\mathcal{X}c_{3}m^2(r-1)\bigr]\bigr\}P_{,X}\nonumber\\
    &&{}+\bigl\{-2(r-1)[(4+\alpha)r-3\alpha-10]P\nonumber\\
    &&{}+\Mpl^2\bigl[[(-30\alpha-120)r^2+(-12\alpha^2-24\alpha-36)r+24\alpha^2+198\alpha+300]H^2\nonumber\\
    &&{}+\mathcal{X}c_{3}m^2(r-1)[(4+\alpha)r-3\alpha-10]\bigr]\bigr\}[\Mpl^2(\mathcal{X}c_{3}m^2-6H^2)-2P](r-1)\Bigr)G_{,X}\nonumber\\
    &&{}+4\Mpl^4H^4(r+1)^2P_{,X}^3\nonumber\\
    &&{}+4H^2\bigl\{(-2r^2+2)P+\Mpl^2\bigl[[(2\alpha+8)r^3+(2\alpha^2+14\alpha+28)r^2+(-2\alpha^2-18\alpha-20)r\nonumber\\
    &&{}-\tfrac52\alpha^2-18\alpha-16]H^2+m^2c_{3}\mathcal{X}(r-1)(r+1)\bigr]\bigr\}\Mpl^2P_{,X}^2\nonumber\\
    &&{}+[\Mpl^2(\mathcal{X}c_{3}m^2-6H^2)-2P](r-1)\bigl\{(-2r+2)P+\Mpl^2\bigl[[(8\alpha+32)r^2+(4\alpha^2+16\alpha+34)r\nonumber\\
    &&{}-8\alpha^2-72\alpha-114]H^2+m^2c_{3}\mathcal{X}(r-1)\bigr]\bigr\}P_{,X}\nonumber\\
    &&{}+2[\Mpl^2(\mathcal{X}c_{3}m^2-6H^2)-2P]^2[(4+\alpha)r-3\alpha-7](r-1)^2\,.
\end{eqnarray}

In the case $\alpha=-4$, we explicitly write, in the following, the no-ghost condition:
\begin{eqnarray}
   Q^{(\alpha=-4)}&=&\frac12\,\frac{\Xi_5^2}{\Xi}\,,\\
  \Xi_5 &=&6-P_{,X}-3\Mpl^{3}\,H^{4}G_{,{\it XX}}-\Mpl^{2}H^{2}P_{,{\it XX}}-12\Mpl\,H^{2}\,G_{,X}\,.
\end{eqnarray}

\section{Cubic horndeski} \label{sec:cubichorndeski}

In this section we present the steps leading to the complete Hamiltonian analysis of a cubic Horndeski theory. Indeed, the construction of the minimal theory relies heavily on the structure already present at this simpler level; for this reason, the reader may find expressions which have already appeared in the main text. We rewrote these to make the appendix as self-contained as possible. To the best of our knowledge, this analysis has never been performed before in the literature.

\subsection{Lagrangian}
The full Lagrangian density of the cubic Horndeski theory is given by 
\begin{equation}
\mathcal{L}_\mathrm{H3}=\mathcal{L}_{\textrm{E-H}}+\mathcal{L}_{\sigma}\,.\label{eq:lagrangian_H3}
\end{equation}
where $\mathcal{L}_{\mathrm{E-H}}= M^2_\mathrm{P}\sqrt{-g} R[g] /2$ is the Einstein-Hilbert Lagrangian density, without cosmological constant. We define the scalar part of the (shift invariant) cubic Horndeski Lagrangian by use of Lagrange multipliers 
\begin{equation}
\mathcal{L}_{\sigma}=\sqrt{-g}\left[F(X,S)+\chi\left(X-\mathfrak{X}\right)+\theta S+g^{\mu\nu}\partial_{\mu}\theta\partial_{\nu}\sigma\right],\label{eq:lagrangian_H3_scalarpart}
\end{equation}
where we write the canonical kinetic term for the scalar field $\sigma$ as
\begin{equation}
\mathfrak{X}\equiv-\frac{1}{2}g^{\mu\nu}\partial_{\mu}\sigma\partial_{\nu}\sigma,\label{eq:H3_canonicalterm}
\end{equation}
and where
\begin{equation}
F(X,S)\equiv P(X)-G(X)S\,,
\end{equation}
in which $P(X)$ and $G(X)$ are sufficiently well-behaved general functions. The Lagrangian density (\ref{eq:lagrangian_H3}) is equivalent to the usual expression of the cubic Horndeski Lagrangian density once the e.o.m.\ of $X$, $\chi$ $\theta$, and $S$ are taken into account. The e.o.m. of $X$, $\chi$ $\theta$, and $S$, calculated from (\ref{eq:lagrangian_H3}), are respectively
\begin{equation}
	\begin{cases}
 	& F_{,X}+\chi=0\,,\\
 	& X-\mathfrak{X}=0\,,\\
 	& S-\Box\sigma=0\,,\\
 	& \theta-G(X)=0\,.
	\end{cases}
\end{equation}
where we have used subscripts after a comma to denote derivatives, for instance, $F_{,X} \equiv \frac{\partial F}{\partial X}$. The system of equations is trivially solved by $\chi=-F_{,X}$, $X=\mathfrak{X}$, $\theta=G(\mathfrak{X})$, and $S=\Box\sigma$, and after replacing this solution in the Lagrangian density (\ref{eq:lagrangian_H3}) one recovers its standard form. By using Lagrange multipliers one can evade all second or higher time derivatives. The equation of motion for the scalar field $\sigma$ is 
\begin{equation}
\chi\Box\sigma + \Box\theta + g^{\mu\nu}\nabla_\mu\chi\nabla_\nu\sigma = 0\,,
\end{equation}
which also reduces to the usual equation of motion once the auxiliary fields have been integrated out,
\begin{equation}
F,_\mathfrak{X}\Box\sigma - G,_\mathfrak{X}\Box(\mathfrak{X}) + g^{\mu\nu}F,_{\mathfrak{X}\mathfrak{X}}\nabla_\mu(\mathfrak{X})\nabla_\nu\sigma = 0\,.
\end{equation}

We use here the (3+1) ADM decomposition of the 4-dimensional metric, which necessitates to define the lapse function $N$, the shift vector $N^i$ as well as the spatial 3-dimensional metric $\gamma_{ij}$. These are defined via the line element
\begin{equation}
ds^2 = -N^2 dt^2 + \gamma_{ij}(N^i dt + dx^i)(N^j dt + dx^j)\,,
\end{equation}
where the indices $i,j,\cdots \in \{1,2,3\}$ are used as spatial indices. The 4-dimensional metric $g_{\mu\nu}$ and its inverse $g^{\mu\nu}$ are then given by 
\begin{gather}
g_{00} = -N^2 + \gamma_{ij} N^i N^j\,,\quad g_{0i} = \gamma_{ij} N^i\,,\quad g_{ij} = \gamma_{ij}\,,\\
g^{00} = -\frac{1}{N^2} \,,\quad g^{0i} = \frac{N^i}{N^2}\,,\quad g^{ij} = \gamma^{ij} - \frac{N^i N^j}{N^2}\,,
\end{gather}
where the spatial indices of the lapse function are raised and lowered using the spatial metric $\gamma_{ij}$, and its inverse $\gamma^{ij}$. In Eq.\ (\ref{eq:H3_canonicalterm}) we have used, and define from here on 
\begin{equation}
\partial_{\perp}*=\frac{1}{N}\left(\dot{*}-N^{i}\partial_{i}*\right)\,
\end{equation}
where $\ast$ stands for any field. Using these definitions we have for example that
\begin{equation}
\mathfrak{X}=\frac{1}{2}\left[(\partial_{\perp}\sigma)^{2}-\gamma^{ij}\partial_{i}\sigma\partial_{j}\sigma\right].
\end{equation}

\subsection{Variables, conjugated momenta, and primary constraints} \label{sec:conjmom}

We consider $\{\gamma_{(ij)},\sigma,X,\chi,\theta,S\}$ and their conjugate momenta as 22 canonical
variables, and $\{N,N^{i}\}$ as Lagrange multipliers, as these only appear linearly in the action. Upon calculating the conjugate momenta, we get 
\begin{gather}
\pi^{ij}\equiv\frac{M_{\mathrm{P}}^{2}}{2}\sqrt{\gamma}\left(K^{ij}-K\gamma^{ij}\right)\,,\quad\pi_{\sigma}\equiv-\sqrt{\gamma}(\chi\partial_{\perp}\sigma+\partial_{\perp}\theta)\,,\quad\pi_{\theta}\equiv-\sqrt{\gamma}(\partial_{\perp}\sigma)\,,\label{eq:momenta_cubichorndeski}\\
\pi_{\chi}=0\,, \quad \pi_{S}=0\,,\quad\pi_{X}=0\,,
\end{gather}
where 
\begin{equation}
K_{ij}=\frac{1}{2N}\left(\dot{\gamma}_{ij}-\mathcal{D}_{i}N_{j}-\mathcal{D}_{j}N_{i}\right)\,.
\end{equation}
As an intermediate step before computing the Hamiltonian, we invert relations (\ref{eq:momenta_cubichorndeski}) as 
\begin{align}
 & \dot{\gamma}_{ij}=2NK_{ij}(\pi^{kl})+\mathcal{D}_{i}N_{j}+\mathcal{D}_{j}N_{i}\\
 & \dot{\sigma}=-N\tilde{\pi}_{\theta}+N^{i}\partial_{i}\sigma\\
 & \dot{\theta}=N(\chi\tilde{\pi}_{\theta}-\tilde{\pi}_{\sigma})+N^{i}\partial_{i}\theta\,.
\end{align}
We have found useful to define the tilded momenta as three dimensional scalars, i.e., for instance, $\tilde{\pi}_\theta \equiv \frac{\pi_\theta}{\sqrt{\gamma}}$. In addition to previous relations, the primary constraints related to $X$, $\chi$, and $S$ are defined as
\begin{equation}
0=P_{X}\equiv\pi_{X}\,,\quad0=P_{\chi}\equiv\pi_{\chi}\,,\quad0=P_{S}\equiv\pi_{S}\,.
\end{equation}

\subsection{Primary Hamiltonian, constraint algebra, and consistency conditions}

The Hamiltonian with all primary constraints can be now written as
\begin{equation}
\bar{H}^{(1)}_\mathrm{H3}=\int d^{3}x\,[-N\mathcal{R}_{0}-N^{i}\mathcal{R}_{i}+\xi_{X}P_{X}+\xi_{\chi}P_{\chi}+\xi_{S}P_{S}]\,,
\end{equation}
where the Lagrange multipliers $\lambda_X$, $\lambda_\chi$ and $\lambda_S$ are scalars (density weight 0) of mass dimension 5, 3, and 4, respectively and where we define
\begin{eqnarray*}
\mathcal{R}_{0} & = & \frac{M_{\mathrm{P}}^{2}}{2}\sqrt{\gamma}\,R[\gamma]-\frac{2}{M_{\mathrm{P}}^{2}}\sqrt{\gamma}\left(\gamma_{il}\gamma_{jk}-\frac{1}{2}\gamma_{ij}\gamma_{kl}\right)\tilde{\pi}^{ij}\tilde{\pi}^{kl}-\mathcal{H}_{\sigma}\,,\\
\mathcal{H}_{\sigma} & = & \sqrt{\gamma}\left[\frac{\chi}{2}\tilde{\pi}_{\theta}^{2}-\tilde{\pi}_{\theta}\tilde{\pi}_{\sigma}-F-\chi\left(X+\frac{1}{2}\gamma^{ij}\partial_{i}\sigma\partial_{j}\sigma\right)-\theta S-\gamma^{ij}\partial_{i}\sigma\partial_{j}\theta\right]\,,\\
\mathcal{R}_{i} & = & 2\sqrt{\gamma}\gamma_{ik}\mathcal{D}_{j}\tilde{\pi}^{kj}-\pi_{\sigma}\partial_{i}\sigma-\pi_{\theta}\partial_{i}\theta\,.
\end{eqnarray*}
Just as in general relativity, the Hamiltonian is vanishing on the constraint surface. For further use we define the equivalent of the scalar canonical term $\mathfrak{X}$ in the Hamiltonian language,
\begin{equation}
\mathfrak{X}_H \equiv \frac{1}{2}\left(\tilde{\pi}_\theta^2 - \gamma^{ij}\partial_i\sigma\partial_j\sigma\right)\,.
\end{equation}

From here one can compute the algebra of the primary constraints.
The relations will be helpful for computing the evolution of the constraints.
First, it is found that if we want to have the momentum constraint
as a generator for translations, we need to modify it so as to include
$\pi_{X}$, $\pi_{\chi}$, and $\pi_{S}$. We thus define 
\begin{align}
\tilde{\mathcal{R}}_{i} & \equiv2\sqrt{\gamma}\gamma_{ik}\mathcal{D}_{j}\tilde{\pi}^{kj}-\pi_{\sigma}\partial_{i}\sigma-\pi_{X}\partial_{i}X-\pi_{\chi}\partial_{i}\chi-\pi_{S}\partial_{i}S-\pi_{\theta}\partial_{i}\theta\,.\label{eq:defmomentumconstrainttilde_H3}
\end{align}
Clearly, this momentum constraint is equivalent to the original one
when restricted to the constraint surface. We turn to analyzing the
constraint algebra, which is summarized in table \ref{table:primaryconstraintalgebra_cubichorndeski}\footnote{In the entries of all tables we have omitted Dirac $\delta$-functions, unless otherwise stated. In particular, we give more details whenever the result of the Poisson brackets formally includes derivatives of $\delta$-functions -- i.e.\ when these cannot be factorized out. See appendix \ref{apx:poissonbrackets} for more details.}. The reader will find in appendix \ref{apx:poissonbrackets} definitions concerning Poisson brackets and their algebra.

\begin{table}[ht]
\begin{tabular}{c|ccccc}
\{$\downarrow$,$\rightarrow$\} & $\mathcal{R}_{0}$  & $\tilde{\mathcal{R}}_{i}$  & $P_{X}$  & $P_{\chi}$  & $P_{S}$ \tabularnewline
\hline 
$\mathcal{R}_{0}$  & $0$  & $0$  & $-(\chi+F_{,X})$  & $-(X-\mathfrak{X}_H)$  & $-(\theta-G(X))$ \tabularnewline 
$\tilde{\mathcal{R}}_{i}$  & \multicolumn{1}{c}{} & $0$  & $0$  & $0$  & $0$ \tabularnewline
$P_{X}$  & \multicolumn{2}{c}{} & $0$  & $0$  & $0$ \tabularnewline 
$P_{\chi}$  & \multicolumn{3}{c}{} & $0$  & $0$ \tabularnewline
$P_{S}$  & \multicolumn{4}{c}{} & $0$ \tabularnewline
\end{tabular}\caption{Primary constraint algebra in the cubic Horndeski theory. Dirac $\delta$-functions were omitted in the entries.}
\label{table:primaryconstraintalgebra_cubichorndeski} 
\end{table}

We give some results in their integral form,
\begin{align}
\{ P_\star,\tilde{\mathcal{R}}_i [f^i] \} & = - \int d^3x \sqrt{\gamma}\,\mathcal{D}_i\!\left(P_\star f^i\right) \approx 0\, ,\label{eq:poisson_primaryWmomentum_H3}\\
\{ \mathcal{R}_0 [\phi_2], \mathcal{R}_0[\phi_2] \} & = \int d^3x \mathcal{R}_i \left(\phi_1\mathcal{D}^i \phi_2-\phi_2\mathcal{D}^i \phi_1\right) \approx 0\,,\\
\{ \mathcal{R}_0 [\phi], \tilde{\mathcal{R}}_i[f^i] \} & = \int d^3x \mathcal{R}_0 f^i\mathcal{D}_i \phi \approx 0\,,\\
\{ \tilde{\mathcal{R}}_i [f^i], \tilde{\mathcal{R}}_j[g^j] \} & = \int d^3x \tilde{\mathcal{R}}_i \left(g^j\mathcal{D}_j f^i -f^j\mathcal{D}_jg^i\right) \approx 0\,.
\end{align}
One can observe that the Hamiltonian and momentum constraints obey the usual algebra. In equation (\ref{eq:poisson_primaryWmomentum_H3}) the symbol $\star$ stands for any of $X$, $\chi$, and $S$. 

Armed by the complete algebra of constraints one can move on to study the consistency conditions. Consistency of the primary constraints $P_X$, $P_\chi$, and $P_S$ with the time evolution of the system yields the following conditions which cannot be solved for Lagrange multipliers (unless one sets N to be zero, which is unphysical): 
\begin{align}
\dot{P}_X  \equiv \sqrt{\gamma}\{\pi_X,\bar{H}_\mathrm{H3}^{(1)}\} & = N \sqrt{\gamma} (\chi+F,_X) \approx 0 \\
\dot{P}_\chi \equiv \sqrt{\gamma}\{\pi_\chi,\bar{H}_\mathrm{H3}^{(1)}\} & = N \sqrt{\gamma} (X-\mathfrak{X}_H) \approx 0 \\
\dot{P}_S   \equiv \sqrt{\gamma}\{\pi_S,\bar{H}_\mathrm{H3}^{(1)}\} & =  N \sqrt{\gamma} (\theta-G(X)) \approx 0 \,.
\end{align}
We thus use these conditions to define the secondary constraints
\begin{eqnarray}
S_X (X, \chi, S) &\equiv& \chi+F,_X\,, \label{eq:constraint_secondary_X_H3}\\
S_\chi (\gamma, \sigma, X, \pi_\theta) &\equiv& X-\mathfrak{X}_H\,,\label{eq:constraint_secondary_chi_H3}\\
S_S (\theta,X) &\equiv& \theta-G(X) \label{eq:constraint_secondary_S_H3}\,.
\end{eqnarray}

\subsection{Secondary Hamiltonian, constraint algebra, and consistency conditions}

We can now define the Hamiltonian with all primary and secondary constraints
\begin{equation}
\bar{H}^{(2)}_\mathrm{H3}=\int d^{3}x\,[-N\mathcal{R}_{0}-N^{i}\tilde{\mathcal{R}}_{i}+\xi_{X}P_{X}+\xi_{\chi}P_{\chi}+\xi_{S}P_{S}+\sqrt{\gamma}\left(\lambda_{X}S_{X}+\lambda_{\chi}S_{\chi}+\lambda_{S}S_{S}\right)]\,,
\end{equation}
where the Lagrange multipliers $\lambda_X$, $\lambda_\chi$, and $\lambda_S$ are spatial scalars, i.e.\ have a density weight of 0, of mass dimension 4, 0, and 3, respectively. 

\begin{table}[ht]
\begin{tabular}{c|cccccccc}
\{$\downarrow$,$\rightarrow$\} & $\mathcal{R}_{0}$  & $\tilde{\mathcal{R}}_{i}$  & $P_{X}$  & $P_{\chi}$  & $P_{S}$  & $S_{X}$  & $S_{\chi}$  & $S_{S}$\tabularnewline
\hline 
$\mathcal{R}_{0}$  & $0$  & $0$  & $0$  & $0$  & $0$  & $0$  & $T_{\chi}$  & $T_{S}$\tabularnewline
$\tilde{\mathcal{R}}_{i}$  & \multicolumn{1}{c}{} & $0$  & $0$  & $0$  & $0$  & $0$  & $0$  & $0$\tabularnewline 
$P_{X}$  & \multicolumn{2}{c}{} & $0$  & $0$  & $0$  & $-F_{,XX}$  & $-1$  & $G_{,X}$\tabularnewline
$P_{\chi}$  & \multicolumn{3}{c}{} & $0$  & $0$  & $-1$  & $0$  & $0$\tabularnewline 
$P_{S}$  & \multicolumn{4}{c}{} & $0$  & $G_{,X}$  & $0$  & $0$\tabularnewline 
$S_{X}$  & \multicolumn{5}{c}{} & $0$  & $0$  & $0$\tabularnewline
$S_{\chi}$  & \multicolumn{6}{c}{} & $0$  & $\pi_{\theta}/\sqrt{\gamma}$\tabularnewline
$S_{S}$  & \multicolumn{7}{c}{} & $0$
\end{tabular}\caption{Secondary constraint algebra in the cubic Horndeski theory. Dirac $\delta$-functions were omitted in the entries.}\label{table_algebrasecondary_H3} 
\end{table}

The secondary constraint algebra is summarized in table \ref{table_algebrasecondary_H3} where $T_{\chi}$ and $T_{S}$ stand for 
\begin{align}
T_{\chi} & =\frac{2}{M_{\textrm{P}}^{2}}\tilde{\pi}^{ij}\left[\gamma_{ij}\mathfrak{X}_H-\mathcal{D}_{i}\sigma\mathcal{D}_{j}\sigma\right]-\tilde{\pi}_{\theta}\left[S-\gamma^{ij}\mathcal{D}_{i}\mathcal{D}_{j}\sigma\right]-\gamma^{ij}\mathcal{D}_{j}\sigma\mathcal{D}_{i}\tilde{\pi}_{\theta}\,,\label{eq:bracket_tertiary_chi_H3}\\
T_{S} & =\chi\tilde{\pi}_{\theta}-\tilde{\pi}_{\sigma}\label{eq:bracket_tertiary_S_H3}\,.
\end{align} 
One can simplify a little the algebra by defining a revised version of the Hamiltonian constraint, 
\begin{equation}
\tilde{\mathcal{R}}_{0}=\mathcal{R}_{0}-T_{\chi}(P_{X}-F_{,XX}P_{\chi})\,,
\end{equation}
which in turn yields Table \ref{table_algebrasecondarybis_H3}.

\begin{table}[ht]
\begin{tabular}{c|cccccccc}
\{$\downarrow$,$\rightarrow$\} & $\tilde{\mathcal{R}}_{0}$  & $\tilde{\mathcal{R}}_{i}$  & $P_{X}$  & $P_{\chi}$  & $P_{S}$  & $S_{X}$  & $S_{\chi}$  & $S_{S}$\tabularnewline
\hline 
$\tilde{\mathcal{R}}_{0}$  & $0$  & $0$  & $0$  & $0$  & $0$  & $0$  & $0$  & $T_{S}+G_{,X}T_{\chi}$\tabularnewline 
$\tilde{\mathcal{R}}_{i}$  & \multicolumn{1}{c}{} & $0$  & $0$  & $0$  & $0$  & $0$  & $0$  & $0$\tabularnewline
$P_{X}$  & \multicolumn{2}{c}{} & $0$  & $0$  & $0$  & $-F_{,XX}$  & $-1$  & $G_{,X}$\tabularnewline
$P_{\chi}$  & \multicolumn{3}{c}{} & $0$  & $0$  & $-1$  & $0$  & $0$\tabularnewline
$P_{S}$  & \multicolumn{4}{c}{} & $0$  & $G_{,X}$  & $0$  & $0$\tabularnewline
$S_{X}$  & \multicolumn{5}{c}{} & $0$  & $0$  & $0$\tabularnewline 
$S_{\chi}$  & \multicolumn{6}{c}{} & $0$  & $\pi_{\theta}/\sqrt{\gamma}$\tabularnewline
$S_{S}$  & \multicolumn{7}{c}{} & $0$
\end{tabular}\caption{Secondary constraint algebra in the cubic Horndeski theory, with modified Hamiltonian constraint. Dirac $\delta$-functions were omitted in the entries.}
\label{table_algebrasecondarybis_H3} 
\end{table}

The consistency conditions yield the following equations 
\begin{align}
\dot{P}_{\chi} \approx0\approx \dot{P}_{S} & :\lambda_{X}\approx0\\
\dot{P}_{X}\approx 0 & :\lambda_{\chi}-\lambda_{S}G_{,X}\approx0\label{consistencyPX_H3}\\
\dot{S}_{\chi}\approx 0 & :\xi_{X}+\lambda_{S}\tilde{\pi}_{\theta}\approx0\label{consistencySchi_H3}\\
\dot{S}_{X}\approx 0 & :\xi_{X}F_{,XX}+\xi_{\chi}-\xi_{S}G_{,X}\approx0\\
\dot{S}_{S}\approx 0 & :N(T_{S}+G_{,X}T_{\chi})+\xi_{X}G_{,X}+\lambda_{\chi}\tilde{\pi}_{\theta}\approx0\label{consistencySS_H3}\\
\dot{\tilde{\mathcal{R}}}_{0}\approx 0 & :\lambda_{S}(T_{S}+G_{,X}T_{\chi})\approx0\,.
\end{align}
By plugging Eqs.\ (\ref{consistencyPX_H3}) and (\ref{consistencySchi_H3}) into
Eq.\ (\ref{consistencySS_H3}), we obtain 
\begin{equation}
N(T_{S}+G_{,X}T_{\chi})=0.
\end{equation}

As a consequence, since setting the lapse to zero would be unphysical,
we need to impose the tertiary constraint $T\approx 0$, where
\begin{equation}
T=T_{S}+G_{,X}T_{\chi}=\chi\tilde{\pi}_{\theta}-\tilde{\pi}_{\sigma}-G_{,X}\left[\frac{2}{M_{\textrm{P}}^{2}}\tilde{\pi}^{ij}\left(\mathcal{D}_{i}\sigma\mathcal{D}_{j}\sigma+\gamma_{ij}\mathfrak{X}_H\right)+\tilde{\pi}_{\theta}\left(S-\gamma^{ij}\mathcal{D}_{i}\mathcal{D}_{j}\sigma\right)+\gamma^{ij}\mathcal{D}_{j}\sigma\mathcal{D}_{i}\tilde{\pi}_{\theta}\right]
\end{equation}

In order to simplify further the constraint algebra, we can form the following combinations

\begin{align}
\tilde{P}_{X} & =P_{X}-F_{,XX}P_{\chi}\,,\\
\tilde{P}_{S} & =\frac{1}{\sqrt{\gamma}}\left(P_{S}+G_{,X}P_{\chi}\right)\,,\\
\tilde{S}_{S} & =S_{S}+G_{,X}S_{\chi}-\tilde{\pi}_{\theta}\tilde{P}_{X}\,,
\end{align}

where $\tilde{P}_{S}$ is a first-class constraint. The resulting algebra is summarized
in table \ref{table:algebratertiary_H3}, where 
\begin{align}
A & =G_{,XX}T_{\chi}+F_{,XX}D\,,\\
D & =\tilde{\pi}_{\theta}\,,\\
Q_{\chi}(x,y) & =\delta(x-y)\left[G_{,X}\frac{1}{M_{\mathrm{P}}^{2}}\left(\tilde{\pi}_{\theta}^{2}\gamma^{ij}-\gamma^{ki}\gamma^{lj}\mathcal{D}_{k}\sigma\mathcal{D}_{l}\sigma\right)\left(\mathcal{D}_{i}\sigma\mathcal{D}_{j}\sigma+\gamma_{ij}\mathfrak{X}_H\right)-\frac{1}{2}\gamma^{ij}\mathcal{D}_{i}\mathcal{D}_{j}\sigma\right]\nonumber \\
 & \quad-\frac{1}{2}\left\{ \left[\gamma^{ij}\mathcal{D}_{i}\sigma\right]\!(y)\,\mathcal{D}_{j}^{(y)}\delta(x-y)-\left[\gamma^{ij}\mathcal{D}_{i}\sigma\right]\!(x)\,\mathcal{D}_{j}^{(x)}\delta(x-y)\right\} \,,\\
B & =P_{,X}+P_{,XX}\tilde{\pi}_{\theta}^{2}-2G_{,X}\gamma^{ij}\mathcal{D}_{i}\mathcal{D}_{j}\sigma-G_{,XX}\left[\tilde{\pi}_{\theta}^{2}\gamma^{ij}\mathcal{D}_{i}\mathcal{D}_{j}\sigma-\gamma^{ij}\gamma^{kl}\mathcal{D}_{i}\mathcal{D}_{k}\sigma\mathcal{D}_{j}\sigma\mathcal{D}_{l}\sigma\right]\nonumber \\
 & +\frac{1}{M_{\mathrm{P}}^{2}}\left[2G_{,X}\tilde{\pi}\tilde{\pi}_{\theta}+G^{2}_{3,X}\left(\frac{3}{2}\tilde{\pi}_{\theta}^{4}-\frac{1}{2}\left(\gamma^{ij}\mathcal{D}_{i}\sigma\mathcal{D}_{j}\sigma\right)^{2}-\tilde{\pi}_{\theta}^{2}\gamma^{ij}\mathcal{D}_{i}\sigma\mathcal{D}_{j}\sigma\right)\right]\nonumber \\
 & +\frac{G_{,XX}}{M_{\mathrm{P}}^{2}}\left[\tilde{\pi}_{\theta}^{2}\tilde{\pi}+\tilde{\pi}_{\theta}\mathcal{D}_{i}\sigma\mathcal{D}_{j}\sigma\left(2\tilde{\pi}^{ij}-\gamma^{ij}\tilde{\pi}\right)\right]\,,
\end{align}
where $A$, $D$, and $B$ are purely local expressions.

\begin{table}[ht]
\begin{tabular}{c|ccccccc}
\{$\downarrow$,$\rightarrow$\} & $\tilde{\mathcal{R}}_{0}$  & $\tilde{P}_{X}$  & $P_{\chi}$  & $S_{X}$  & $S_{\chi}$  & $\tilde{S}_{S}$ & $T$\tabularnewline
\hline 
$\tilde{\mathcal{R}}_{0}$  & $0$  & $0$  & $0$  & $0$  & $0$  & $0$  & $-\tilde{Q}$\tabularnewline
$\tilde{P}_{X}$  & \multicolumn{1}{c}{} & $0$  & $0$  & $0$  & $-1$  & $0$  & $A$\tabularnewline
$P_{\chi}$  & \multicolumn{2}{c}{} & $0$  & $-1$  & $0$  & $0$  & $D$\tabularnewline
$S_{X}$  & \multicolumn{3}{c}{} & $0$  & $0$  & $0$  & $0$\tabularnewline
$S_{\chi}$ & \multicolumn{4}{c}{} & $0$  & $0$  & $-Q_{\chi}$\tabularnewline
$\tilde{S}_{S}$  & \multicolumn{5}{c}{} & $0$  & $B$\tabularnewline
$T$  & \multicolumn{6}{c}{} & $Q_{T}$
\end{tabular}\caption{Tertiary constraint algebra, with some new combinations, in the cubic Horndeski thoery. First-class constraints were omitted.  Dirac $\delta$-functions were omitted in the entries, excepting for $\tilde{Q}$, $Q_\chi$, and $Q_T$, which yield derivatives of $\delta$-functions.}
\label{table:algebratertiary_H3} 
\end{table}

At this point we define a new Hamiltonian constraint that will commute
with all the constraints. Such a constraint becomes thus first-class
in the tertiary constraint algebra. 
\begin{equation}
\bar{\mathcal{R}}_{0}=\tilde{\mathcal{R}}_{0}+\frac{\tilde{Q}}{B}\tilde{S}_{S}
\end{equation}
This combination is well defined under the condition that $B\neq0$. In the case $B=0$, the constraint $\tilde{S}_S$ becomes first-class. As a result, we have the right number of constraints even though the Hamiltonian constraint is not first-class. In what follows we will assume that $B\neq0$.

\subsection{Closedness of the algebra and total Hamiltonian}

To show that the algebra is closed, we need to show that 
\begin{equation}
\det\left(\mathcal{M}(x,y)\right)\neq0\,, \label{eq:closedness}
\end{equation}
where $\mathcal{M}(x,y)$ is the constraint algebra matrix. This is
equivalent to showing that, for all $\vec{u}$, 
\begin{equation}
\int d^{3}xd^{3}y\vec{u}^{\intercal}(x)\mathcal{M}(x,y)\vec{v}(y)=0\,,
\end{equation}
implies that $\vec{v}=0$.

We make the constraint algebra explicit, and obtain 
\begin{align}
0=\int d^{3}xd^{3}y & \left\{ u_{1}\left[-v_{5}+Av_{6}\right]+u_{2}\left[-v_{3}-\tilde{\pi}_{\theta}v_{6}\right]+u_{3}\left[v_{2}\right]+u_{4}\left[Bv_{6}\right]+u_{5}\left[v_{1}-Q_{\chi}^{0}v_{6}-Q_{\chi}^{i}\partial_{i}v_{6}\right]\right.\nonumber \\
 & \left.u_{6}\left[-Av_{1}+\tilde{\pi}_{\theta}v_{2}-Bv_{4}+Q_{\chi}^{0}v_{5}+Q_{\chi}^{i}\partial_{i}v_{5}-\partial_{i}\left(Q_{T}^{i}v_{6}\right)-Q_{T}^{i}\partial_{i}v_{6}\right]\right\} \,.
\end{align}
Here we have defined $Q_\chi^0$, $Q_\chi^i$, $Q_T^0$, and $Q_T^i$ by
\begin{align}
\iint d^3x d^3y f_1(x) Q_\chi(x,y) f_2(y) &= \int d^3 x f_1 \left(Q_\chi^0 f_2 + Q_\chi^i\partial_if_2\right)\,,\\
\iint d^3x d^3y f_1(x) Q_T(x,y) f_2(y) &= \int d^3 x \left[f_1 f_2 Q_T^0 + Q_T^i\left(f_2\partial_if_1-f_1\partial_if_2\right)\right]\,,
\end{align}
where $f_1$ and $f_2$ are any auxiliary well-behaved functions. As a consequence of setting the whole integral to zero we need to impose each expression between square brackets to vanish separately. As $B\neq 0$ we can use the first 5 square brackets to say $v_1=v_2=v_3=v_5=v_6=0$. By plugging these into the last squared bracket we obtain that also $v_4=0$ necessarily. The determinant of $\mathcal{M}(x,y)$ is thus different from zero.

The candidates for the second-class constraints are therefore indeed second-class if 
$B\neq0$. In such a case the candidates for the first-class constraints are indeed first-class. The algebra is then closed. Furthermore, as the Hamiltonian is only composed of constraints, we have shown that the tertiary Hamiltonian is the total Hamiltonian and that there are no further constraints given by the consistency conditions.

\subsection{Tertiary Hamiltonian}

The total Hamiltonian of the cubic Horndeski theory is given by 
\begin{equation}
\bar{H}^{(T)}_\mathrm{H3}=\int d^{3}x\,[-N\bar{\mathcal{R}}_{0}-N^{i}\tilde{\mathcal{R}}_{i}+\xi_{X}\tilde{P}_{X}+\xi_{\chi}P_{\chi}+\sqrt{\gamma}\,(\xi_{S}\tilde{P}_{S}+\lambda_{X}S_{X}+\lambda_{\chi}S_{\chi}+\lambda_{S}\tilde{S}_{S}+\lambda_T T)]\,.
\end{equation}
where the total number of degrees of freedom is 3. This is due to the presence of the 5 first-class constraints ($\bar{\mathcal{R}}_0, \tilde{\mathcal{R}}_i,{\tilde P}_S$) and 6 second-class constraints (the remaining ones), which kill the degrees of freedom introduced by the auxiliary fields.

\section{Poisson brackets}
\label{apx:poissonbrackets}
In order to handle Poisson brackets resulting formally into derivatives of Dirac $\delta$-functions -- in particular when spatial derivatives of the fields and conjugated momenta are present -- we rely on the use of smearings. One can indeed turn any local expression $A(x)$ into a functional by convolution, i.e.
\begin{equation}
A[\phi] = \int d^3 x A(x) \phi(x)\,,
\end{equation}
where $\phi(x)$ is an auxiliary scalar function with well-behaved properties. If needed, we multiply or divide by the number of factors $\sqrt{\gamma}$ needed to turn the integral into a scalar. The Poisson brackets can then be computed as
\begin{equation}
\{A_1[\phi_1],A_2[\phi_2]\} \equiv \int d^3 z \sum_\alpha \left(\frac{\delta A_1[\phi_1]}{\delta q_\alpha(z)}\frac{\delta A_2[\phi_2]}{\delta p_\alpha(z)}-\frac{\delta A_1[\phi_1]}{\delta p_\alpha(z)}\frac{\delta A_2[\phi_2]}{\delta q_\alpha(z)}\right)
\end{equation}
where $\alpha$ runs over all canonical variables $q_\alpha$, which have the $p_\alpha$ as conjugated momenta. We consider that the result of the Poisson bracket, $A_3(x,y)$, is the kernel
\begin{equation}
\{A_1[\phi_1],A_2[\phi_2]\} \approx \iint d^3 x\, d^3 y\;\sqrt{\gamma}^{(x)}\sqrt{\gamma}^{(y)} \phi_1(x) A_3(x,y) \phi_2 (y)
\end{equation}
In most cases, $A_3(x,y)$ is simply proportional to a Dirac $\delta$-function, in which case we can define $A_3 (x)$ as
\begin{equation}
A_3(x,y) = \delta(x-y) A_3 (x)\,.
\end{equation}
However, in some cases, we cannot give such an expression. Throughout the text, in the tables summarizing the constraint algebra we will give the kernel of the resulting expression, omitting the Dirac $\delta$-functions unless otherwise stated.

\end{document}